\documentclass[a4paper,11pt]{article}
\pdfoutput=1 

\usepackage{jcappub} 

\usepackage[T1]{fontenc} 
\usepackage{pifont}
\usepackage[utf8]{inputenc}
\usepackage{booktabs}

\usepackage{fontawesome5}
\usepackage{hyperref}
\usepackage[nameinlink,noabbrev]{cleveref}

\crefname{equation}{Eq.}{Eqs.}

\crefname{figure}{Figure}{Figures}
\crefname{table}{Table}{Tables}
\crefname{appendix}{Appendix}{Appendices}
\Crefname{figure}{Figure}{Figures}
\Crefname{equation}{Equation}{Equations}
\Crefname{section}{Section}{Sections}
\Crefname{table}{Table}{Tables}

\newcommand{\absum}[0]{\textsc{AbacusSummit}}

\title{Exploring non-Poisson satellite occupation in HOD models and its impact on 2- and 3-point galaxy clustering}

\author[a]{Antoine Rocher}

\affiliation[a]{Institute of Physics, Laboratory of Astrophysics, \'Ecole Polytechnique F\'ed\'erale de Lausanne (EPFL), Observatoire de Sauverny, CH-1290 Versoix, Switzerland \\}

\emailAdd{antoine.rocher@epfl.ch}

\abstract{
Understanding the connection between galaxies and dark matter halos is a central challenge in modern cosmology. The Halo Occupation Distribution (HOD) framework provides a widely used statistical description of how galaxies populate dark matter halos, enabling precise modelling of galaxy clustering. A common assumption in standard HOD models is that the number of satellite galaxies follows a Poisson distribution at fixed halo mass. In this work, we revisit this assumption and introduce the Conway–Maxwell–Poisson (CMP) distribution as a minimal extension of of the Poisson model, which add a single parameter, $\nu$, to explore sub- and super-Poisson behaviour. We derive analytical approximations for the CMP expectation parameter $\lambda$ and develop a numerical scheme that smoothly connects small- and large-$\lambda$ regimes, achieving $\sim5\%$ accuracy for $0.55 < \nu < 4$. Using the \texttt{HODDIES} package, we study the impact of non-Poisson satellite occupations on mock galaxy catalogues and clustering statistics. Variations in the variance of the satellite occupation significantly affect small-scale clustering, producing deviations in the in the 2-point function up to $15\%$ for the projected clustering and $10\%$ for the monopole and quadrupole. We further investigate higher-order statistics using counts-in-cylinders (CIC) and the tree-level galaxy bispectrum. CIC statistics are highly sensitive to changes in the variance, with variations up to $\sim40\%$, while the tree-level galaxy bispectrum (in the Sugiyama basis) is only weakly affected, $<4\%$ up to $k_\mathrm{max}=0.3\,h\,\mathrm{Mpc}^{-1}$. These results suggest that non-Poisson satellite statistics are important for small-scale analyses, but should have a limited impact on cosmological constraints from power spectrum and bispectrum measurements using large scales $k_\mathrm{max} < 0.3$.

}

\begin{document}
\maketitle
\flushbottom

\section{Introduction}

The large-scale structure (LSS) of the universe is one of the most powerful tools to study cosmology. Over the last two decades, the field of LSS has evolved from theory to data-driven, and has recently shifted towards Stage-IV cosmological surveys, thanks to experiments such as the Dark Energy Spectroscopic Instrument (DESI,~\cite{abareshi_overview_2022}), the 4-Metre Multi-Object Spectroscopic Telescope (4MOST)~\cite{de_jong_4most_2019}, \textit{Euclid}~\cite{collaboration_euclid_2025} or the upcoming Legacy Survey of Space and Time (LSST;~\cite{ivezic_lsst_2019}). In the coming years, unprecedented improvements in the precision of galaxy clustering measurements will provide opportunities to study the cosmological model and expand our understanding of galaxy formation. Recent results of the DESI experiment combine with other cosmological probes already suggest an evolution of dark energy with time~\cite{collaboration_desi_2025, collaboration_desi_2025-2}, from the measurement of the baryon acoustic oscillation (BAO;\cite{collaboration_desi_2025-4}) scale and the full shape analyses of the two-point correlation function at large-scales~\cite{adame_desi_2025}. While analytical models such as effective field theory (see~\cite{cabass_snowmass_2022, senatore_bias_2015}, for a review) models are efficient to probe the non-linear dark matter structure growth and the galaxy clustering up to quasi-linear scales $>30 [\mathrm{Mpc}/h]$, they break down in smaller, non-linear scales. Since galaxies form and evolve in dark-matter halos~\cite{white_core_1978}, small scales provide valuable information about the connection between galaxies and their host dark matter halos. These scales contribute significantly to constraints on the growth rate of structure, therefore a correct modelling of the galaxy-halo connection is a critical ingredient to extract cosmological information and understand galaxy formation from observed galaxy clustering statistics. In addition, accurately modelling the galaxy-halos connection is essential for the construction of realistic mock catalogues that are used to design large-scale analyses and to evaluate systematic uncertainties~\cite{findlay_exploring_2025, garcia-quintero_hod-dependent_2025}. With the increase of precision of galaxy clustering measurements, uncertainties related to the galaxy--halo connection are expected to become increasingly more relevant and a precise understanding of the galaxy--halo connection will therefore be necessary to derive robust systematic error budgets for cosmological measurements. 

The galaxy--halo connection can be modelled through several complementary approaches, ranging from full-physics hydrodynamical simulations to empirical or physically motivated methods built on top of gravity-only $N$-body simulations (see~\cite{wechsler_connection_2018} for a review). Semi-analytic models (SAMs) describe galaxy formation through a set of coupled differential equations that encode the main physical processes driving galaxy evolution~\cite{cole_hierarchical_2000, baugh_primer_2006, bower_breaking_2006, guo_galaxy_2013, lu_semi-analytic_2014}. Alternatively, subhalo abundance matching (SHAM) provides an empirical framework by assigning galaxies to halos or subhalos through a monotonic relation between galaxy properties (e.g., luminosity or stellar mass) and halo properties, typically with some scatter~\cite{kravtsov_dark_2004, tasitsiomi_modeling_2004, vale_non-parametric_2006, conroy_modeling_2006,favole_clustering_2016}. Another, very popular approach is the Halo Occupation Distribution (HOD)~\cite{berlind_halo_2002, zheng_theoretical_2005} that provides a simple and flexible statistical description of how galaxies populate dark matter halos in N-body simulations. Within this framework, the galaxy population of a halo of mass $M$ is described through the conditional probability distribution $P(N_{\rm gal}\,|\,M)$, typically decomposed into contributions from central and satellite galaxies. The functional form of the HOD depends on the galaxy type. The model proposed in~\cite{zheng_theoretical_2005} has demonstrated considerable success in reproducing the occupation statistics of galaxies selected above a minimum luminosity or stellar mass like Luminous Red Galaxies (LRGs)~\cite{zheng_galaxy_2007, zehavi_galaxy_2011, hadzhiyska_millenniumtng_2023, yuan_desi_2023}. 
In this model the average number of central galaxies $\langle N_\mathrm{c}\rangle$ is well described by a smooth step function while the average number of satellite galaxies follows a power law. For star forming galaxies, such as Emission Line Galaxies (ELGs), the halo occupation for central galaxies is different from a step function and best described by an asymmetric Gaussian distribution~\cite{geach_clustering_2012, gonzalez-perez_host_2018, hadzhiyska_millenniumtng_2023, rocher_desi_2023, avila_completed_2020}. The central galaxies are defined to reside at the centre of mass of the halo and are usually modelled through a Bernoulli process (either 0 or 1 galaxy at the centre of the halo). In contrast, satellite galaxies will reside in the gravitational potential of the DM halo defined by their DM profile, assuming generally a Navarro-Frenk-White profile~\cite{navarro_universal_1997}. The number of satellite galaxies at fixed halo mass is commonly assumed to follow a Poisson distribution with mean $\langle N_{\rm sat}(M) \rangle$. 

Due to its simplicity, the Poisson assumption is not guaranteed to hold in realistic galaxy populations. Several numerical studies have found evidence for departures from Poisson statistics in the abundance of subhaloes within dark matter halos. For example,~\cite{boylan-kolchin_theres_2010} showed that the number of low-mass subhaloes in host halos is better described by a negative binomial distribution, corresponding to a super-Poissonian scatter in the subhalo population. Using semi-analytic galaxy formation models,~\cite{jiang_statistics_2017} further demonstrated that ignoring non-Poisson behaviour in the halo occupation of subhaloes can lead to systematic errors in predicted galaxy clustering. More recently, analyses based on hydrodynamical simulations have also found evidence for deviations from Poisson statistics in satellite galaxy populations. For instance, studies of LRG and ELG galaxy samples constructed from the IllustrisTNG~\cite{nelson_illustristng_2021} simulation indicate that satellite counts can exhibit super-Poissonian scatter in specific halo populations, particularly for halos with low concentration, leading to discrepancies between standard HOD predictions and simulated galaxy--galaxy lensing signals~\cite{hadzhiyska_millenniumtng_2023, chaves-montero_galaxy_2023}. These results suggest that the Poisson assumption commonly adopted in HOD models may not provide a complete description of satellite galaxy statistics. The signature of non-Poisson statistic in the satellite occupation will affect higher-order clustering, which are currently included in main analysis of galaxy clustering~\cite{chudaykin_reanalyzing_2025,damico_boss_2024, lu_preference_2025}. Therefore, allowing for controlled deviations from Poisson statistics offers a promising route to improve the realism of galaxy--halo connection models and to better capture the statistical properties of satellite populations. In former work, several extensions to the Poisson distribution has been studied~\cite{avila_completed_2020, jimenez_extensions_2019, vos-gines_improving_2024} using nearest integer and negative binomial distribution for sub-Poisson and super-Poisson distribution, respectively. 
Ref.~\cite{vos-gines_improving_2024} introduced a framework that allows continuous exploration of sub- and super-Poisson satellite occupations using a single dispersion parameter, $\omega_\sigma$, by combining extended-binomial, Poisson, and negative-binomial distributions, with the nearest-integer distribution providing the minimum-variance limit. This work demonstrated the sensitivity of small-scale two-point clustering and counts-in-cells to the satellite occupation variance. However, this model can lead to negative variance when $\omega < -1/\langle N_\mathrm{sat}\rangle$ where corrections are required. In this work, we investigate the Conway–Maxwell–Poisson (CMP;~\cite{noauthor_r_nodate}) distribution as a complementary prescription. CMP describes sub-Poisson, Poisson, and super-Poisson occupations within a single distribution probability, with one additional dispersion parameter $\nu$ and non-negative probabilities by construction. It also provide a different relation between the variance and mean occupation compare to the formalism of~\cite{vos-gines_improving_2024}. The aim of this paper is to provide an alternative prescription to test for deviation from Poisson statistics using the CMP distribution. We present how the variance of the satellite occupation affects the clustering signal in the two-point correlation function and Count-in-Cylinder (CIC) statistics~\cite{reid_constraining_2009} providing comparison to~\cite{vos-gines_improving_2024}. CIC are an alternative way to probe higher-moment of the galaxy distribution on small scales. In addition, we extend this test to three point function using the tree-level galaxy bispectrum (in the Sugiyama basis~\cite{sugiyama_complete_2019}).


The paper is organized as follows. \Cref{sec:halo occupation statistics} introduce the halo occupation distribution (HOD) framework and discuss the standard Poisson assumption for satellite galaxies, along with motivations for considering deviations from this model. In~\Cref{sec:CMP distribution}, we present CMP distribution and detail the numerical implementation of the CMP distribution, including the evaluation of the normalization constant and sampling procedure. We derive analytical approximations for the mean and variance of the CMP distribution in both the small- and large-$\lambda$ regimes, and construct a smooth interpolation between these limits to obtain an accurate mapping between the target mean occupation number and the expectation parameter of the CMP distribution. \Cref{sec:Impact on mocks} investigates the impact of CMP statistic on HOD-based mock galaxy catalogues and two-point clustering measurements. We describe the impact of non-Poisson statistics on satellite galaxy occupation and the modifications to the one-halo term. We extended this analysis to higher-order statistics using counts-in-cylinders (CIC) and bispectrum measurements to assess their sensitivity to the non-Poisson satellite occupation. Finally, we summarize our results and discuss future prospects in~\Cref{sec:conclusion}.

\section{Halo occupation statistics}
\label{sec:halo occupation statistics}
The Halo Occupation Distribution (HOD) framework provides a powerful statistical description to populate galaxies in dark matter haloes in $N$-body simulation. In the HOD framework, the galaxy population is specified through the conditional probability $P(N_{\mathrm{gal}} \mid M_h)$ which describes the number of galaxies residing in a halo of mass $M_h$. The total occupation is typically decomposed into contributions from central and satellite galaxies $N_{\mathrm{gal}} = N_{\mathrm{cen}} + N_{\mathrm{sat}}$. In standard HOD formalism, central galaxies are modelled as a Bernoulli process, meaning that either zero or one central galaxy will populate a halo. The number of satellite galaxies is assumed to follow a Poisson distribution around a mean $\lambda=\langle N_\mathrm{sat} \rangle$:
\begin{equation}
\label{eq:PMF poisson}
P(k) = \frac{\lambda^k e^{-\lambda}}{k!}.
\end{equation}
The mean of the Poisson distribution is expressed as:
\begin{equation}
\label{eq:mean poisson}
\mathbb{E}(X)=\sum_{k=0}^{\infty} k \frac{\lambda^k e^{-\lambda}}{k!}=\lambda,
\end{equation}
and the variance of the Poisson distribution is equal to its mean,
\begin{equation}
\mathrm{Var}(N_{\mathrm{sat}}\mid M) = \langle N_{\mathrm{sat}}(M)\rangle.
\end{equation}
The Poisson assumption is motivated by the interpretation of satellites as independent tracers within an halo, arising from the superposition of independent merger events. Under this assumption, the second factorial moment satisfies
\begin{equation}
\langle N_{\rm sat}(N_{\rm sat}-1)\rangle = \langle N_{\rm sat}\rangle^2,
\end{equation}
which directly controls the amplitude of the one-halo contribution to clustering, as $\langle N_{\rm sat}(N_{\rm sat}-1)\rangle$ directly connects to the number of satellites-satellites pairs. However, numerical simulations and observational analyses have suggested that satellite occupation may deviate from strict Poisson behaviour, particularly in regimes of high halo mass~\cite{hadzhiyska_millenniumtng_2023}. 
Sub-Poissonian or super-Poissonian fluctuations can arise from halo assembly bias, tidal stripping, correlated infall histories, or baryonic processes affecting satellite survival~\cite{chaves-montero_galaxy_2023}. These effects lead to a modified second moment of the occupation number $\langle N_{\mathrm{sat}}(N_{\mathrm{sat}}-1)\rangle \neq \langle N_{\mathrm{sat}}\rangle^2$, and therefore alter the small-scale clustering signal. Several extensions to the Poisson assumption have been proposed in the literature, including binomial distribution or nearest-integer models, as well as empirical rescaling of the satellite variance~\cite{vos-gines_improving_2024, avila_completed_2020, hadzhiyska_millenniumtng_2023}. Detecting such deviations is challenging, as their impact on two-point clustering statistics is relatively small and can be degenerate with other HOD parameters. The use of high-order statistics like Count-in-Cylinder (CIC) has more constraining power over the variance of the satellite distribution and help to break these degeneracies~\cite{vos-gines_improving_2024}.
In the following, we present an alternative way to study the Poisson behaviour of satellite galaxy occupation.

\section{Describing satellite occupation with the Conway--Maxwell--Poisson distribution}
\label{sec:CMP distribution}
\subsection{Conway--Maxwell--Poisson as a minimal extension}

In this section, we present the Conway--Maxwell--Poisson (CMP;~\cite{noauthor_r_nodate}) distribution as a minimal generalization of the Poisson model for satellite occupation. The CMP distribution introduces a single dispersion parameter $\nu$ that controls deviations from Poisson behaviour while preserving the discrete count nature of the occupation statistics:
\begin{equation}
\label{eq: PMF CMP}
P(N_{\mathrm{sat}} = k \mid \lambda, \nu)
=
\frac{\lambda^k}{(k!)^{\nu} Z(\lambda,\nu)}.
\end{equation}
where $Z(\lambda,\nu)$ is a normalisation constant defined as:
\begin{equation}
\label{eq:normalisation constant}
Z(\lambda,\nu) = \sum_{j=0}^{\infty} \frac{\lambda^j}{(j!)^\nu}
\end{equation}
In this parametrisation the Poisson distribution is recovered when $\nu=1$. The additional parameter $\nu$ controls the dispersion of the distribution:
\begin{equation*}
\nu < 1 \Rightarrow \text{super-Poissonian},
\qquad
\nu > 1 \Rightarrow \text{sub-Poissonian}.
\end{equation*}
The mean and the variance of the CMP distribution are defined as:
\begin{equation}
\begin{aligned}
\label{eq: mean var CMP}
\mathbb{E}[X] &=\lambda \frac{\partial}{\partial \lambda} \log Z(\lambda, \nu) \\
\operatorname{Var}(X)&=\lambda \frac{\partial}{\partial \lambda} E[X].
\end{aligned}
\end{equation}
For computational efficiency the probability mass function in \Cref{eq: PMF CMP} is evaluated in log space:

\begin{equation}
    \log(P(k \mid \lambda, \nu) = k\log{\lambda -\nu\log(k!) - \log(Z(\lambda,\nu))}, \\
\end{equation}
To evaluate the $\log(Z(\lambda,\nu))$ in a numerically stable way, we use the log--sum--exp identity:
\begin{equation}
\begin{aligned}
&\log\left(\sum_j e^{f(j, \lambda)}\right) = m + \log\left(\sum_j e^{f(j, \lambda) - m}\right),\\
& \mathrm{with} \quad m = \max_j f(j, \lambda), \quad f(j, \lambda) = j\log(\lambda) - \nu\log(j!)
\end{aligned}
\end{equation}
Factoring out the largest exponent ensures that all exponential terms satisfy $e^{f(j,\lambda)-m} \le 1$, thereby preventing overflow and underflow in floating--point arithmetic. 
To compute the probabilities $P(k \mid \lambda, \nu)$, the normalizing constant, $Z(\lambda, \nu) $, needs to be determined. The normalizing constant has no closed form but can be approximated numerically by computing the sum up to a large enough value of $j$ that tails contribute negligibly. The minimum number of $j$ iteration that ensure a convergence of the normalizing constant $Z(\lambda, \nu)$ will depend on the value of $\lambda$, with higher number of iterations with increasing $\lambda$. Once the probabilities $P(k \mid \lambda, \nu)$ are computed, the cumulative distribution function (CDF) is obtained by taking the cumulative sum of the probability mass function up to a truncation value $\mathrm{max}_k$. The resulting CDF is then normalized by its final value to ensure that it ranges between 0 and 1. Random samples can subsequently be generated using inverse transform sampling by drawing uniform random numbers in the interval $[0,1]$ and mapping them to integer values through the inverse CDF. The choice of $\mathrm{max}_k$ depends on the value of $\lambda$, since the support of the distribution broadens as $\lambda$ increases. The truncation value max$_k$ was empirically determined. We implement a scaling relation between $\mathrm{max}_k$ and $\lambda$ as follow:
\begin{equation}
\mathrm{max}_k(\lambda) =
\begin{cases}
50, & \lambda < 15, \\
200, & 15 \leq \lambda < 100, \\
1000, & 100 \leq \lambda < 700, \\
5000, & 700 \leq \lambda < 4000, \\
\left\lfloor 1.1\,\lambda \right\rfloor, & \lambda \geq 4000.
\end{cases}
\end{equation}
This scaling ensures that $\mathrm{max}_k$ is always large enough to capture the full support of the distribution without truncating the cumulative probability.  
In a similar way, the determination of the number of iterations $j$ to compute the normalizing constant $Z(\lambda, \nu)$ is set to be $j= 1.1 \cdot \mathrm{max}_k$, with a minimum number of iteration of 300 to always ensure the convergence. This scaling has been tested up to $\lambda = 10^5$, which is largely sufficient for $\langle N_{\rm sat}(M)\rangle$ values.
This design serves two purposes. First, it guarantees that the CMP distribution can be sampled accurately for any value of $\lambda$ without introducing truncation bias. Second, it reduces the number of iterations required for small $\lambda$ values compared to using a fixed large truncation limit, thereby significantly improving the computational efficiency of the sampling procedure.

In the context of HOD, this parametrisation provides a direct way to test the Poisson nature of satellite galaxy statistics. However, contrary to the Poisson case, the CMP distribution the parameter $\lambda$ cannot be directly interpreted as the mean number of satellites. Instead, the mean depends on both $\lambda$ and the dispersion parameter $\nu$ (see \Cref{eq: mean var CMP}). In order to preserve the physical interpretation of the HOD parameter $\langle N_{\rm sat}(M)\rangle$, one must therefore determine a mapping between the target mean occupation and the corresponding value of $\lambda$ for a given $\nu$. In the following, we construct corrections that ensure the resulting CMP distribution preserves the desired mean satellite occupation $\langle N_{\rm sat}(M)\rangle$.


\subsection{Mean and variance scaling across $\lambda$ values}

The primary analytical difficulty lies in the normalizing constant $Z(\lambda,\nu)$, which has no closed form. As a consequence, the mean and variance must either be computed via infinite summation or approximated asymptotically. We developed different approximations depending on the value of $\lambda$ as the structure of the CMP distribution changes qualitatively depending on its value: 
\begin{itemize}
    \item[\ding{227}] For $\lambda << 1$, small values of $k=0,1,2...$ dominates the probability mass.
    \item[\ding{227}] For large $\lambda$, the distribution becomes sharply peaked around specific value of $k$ and admits a saddle-point expansion.
\end{itemize}

\subsection{Regime I: Small $\lambda$}
\label{sec:small regime}
For $\lambda \to 0$, the probability mass of the CMP distribution is concentrated almost on the first few integers $k=0,1,2,\dots$. Truncating the expansion at fourth order in $\lambda$ yields:
\begin{equation}
Z = 1 + \lambda + \frac{\lambda^2}{2^\nu} + \frac{\lambda^3}{6^\nu} + \frac{\lambda^4}{24^\nu} + \mathcal{O}(\lambda^5).
\end{equation}
Using \Cref{eq: mean var CMP} the mean and the variance can be approximate as:
\begin{equation}
\begin{aligned}
\label{eq:mean var CMP approx small lambda}
    \mathbb{E}[X]
&=
\lambda + \lambda^2 \left(2^{1-\nu} - 1\right) + \lambda^3 \left(\frac{3}{6^{\nu}} - \frac{3}{2^{\nu}} + 1\right)\\
&+ \lambda^4 \left(\frac{4}{24^{\nu}} - \frac{4}{6^{\nu}} - \frac{2}{2^{2\nu}} + \frac{4}{2^{\nu}} - 1\right) + \mathcal{O}(\lambda^5)\\
\\
    \mathrm{Var}[X] &= \lambda + \lambda^2\!\left(2^{2-\nu}-2\right) + \mathcal{O}(\lambda^3).
\end{aligned}
\end{equation}
The mean is evaluated up to $\mathcal{O}^4$ as it provide non-negligible correction for $\lambda\sim0.1$ (see \Cref{fig: mean lambda both}). At leading order for the case of $\lambda \to 0$ we get:
\begin{equation}
\mathbb{E}[X] \sim \lambda, \qquad \mathrm{Var}(X) \sim \lambda
\end{equation}
Hence, independently of $\nu$, the CMP distribution is locally indistinguishable from a Poisson distribution at sufficiently small value of $\lambda$. To estimate the $\lambda$ that leads to a mean of the CMP distribution $\mathbb{E} = \langle N_\mathrm{sat}\rangle$ one has to seek a power series given by:
\begin{equation}
\lambda = \mathbb{E} + \alpha  \mathbb{E}^2 + \beta \mathbb{E}^3 +\gamma  \mathbb{E}^4 +\mathcal{O}( \mathbb{E}^5).
\end{equation}
By substituting this power series with the expansion from \Cref{eq:mean var CMP approx small lambda}, the coefficients can be obtained by matching powers of $\mu$ order by order. This procedure yields the inverse relation:
\begin{equation}
\begin{aligned}
\label{eq:mean to lambda small}
\lambda_\mathrm{\to 0}
&=
 \mathbb{E}
- A\, \mathbb{E}^2 
+ (2A^2 - B)\, \mathbb{E}^3 \\
&+ (-5A^3 + 5AB - C)\, \mathbb{E}^4
+ \mathcal{O}( \mathbb{E}^5)
\end{aligned}
\end{equation}
where the coefficients $A,B,C$ depend come from the mean expansion in \Cref{eq:mean var CMP approx small lambda}:
\begin{equation}
\begin{aligned}
    A = 2^{1-\nu}-1,\qquad B = \frac{3}{6^{\nu}}- \frac{3}{2^{\nu}}+1, \\
C = \frac{4}{24^{\nu}} -\frac{4}{6^{\nu}} -\frac{2}{2^{2\nu}} +\frac{4}{2^{\nu}} -1.
\end{aligned}
\end{equation}
Using the above equations we can estimate the correct values of the expectation parameter $\lambda$ that reproduces the excepted mean value $\mathbb{E} = \langle N_{\rm sat}(M)\rangle$. To test this mapping $\langle N_{\rm sat}(M)\rangle$ values from 0.001 to 10 are converted into the CMP parameter $\lambda$ using \cref{eq:mean to lambda small}. Then, we draw $10^6$ samples from the CMP distribution using this aforementioned $\lambda$ value and compare to the targeted mean value for $0.5<\nu<4.5$. The results are shown in the left panel of \Cref{fig: mean lambda both}. We compare the second order $\mathcal{O}(\mathbb{E}^2)$ to the forth order the second order $\mathcal{O}(\mathbb{E}^4)$ approximation. The shaded blue region represent 5\% agreement which is used to define the validity of the approximation. 
The choice of 5\% accuracy is based on the following arguments. First only a few specific ($N_{sat}$, $\nu$) couple are reaching 5\%, otherwise the effect is mainly of 1 to 2\% and results from mocks only lead to deviation $< 1\%$ on the number of satellites galaxies relative to the Poisson case (see \Cref{sec:Impact on mocks and occupation numbers}).
Then, the intrinsic variation in the number of galaxies due to stochastic noise create variation sometimes larger than 5\% in the number of galaxy (especially samples with low density) for a single HOD realisation. Furthermore, HOD studies often allow $\sim10\%$ uncertainties on the galaxy number density when performing HOD fitting against data~\cite{yuan_desi_2023, yuan_span_2022}.
\\
Using the forth order allows to push the validity of the approximation to $\langle N_{\rm sat}(M)\rangle \sim 0.9$ while the second order approximation breaks at $\langle N_{\rm sat}(M)\rangle \sim 0.1$. Using tabulated values we checked that decreasing to $\nu=0.5$ leads to a deviation of 6.4\% and increasing to $\nu=4.5$ leads to a 7.2\% deviation up to $\langle N_{\rm sat}(M)\rangle = 0.9$. For higher $\langle N_{\rm sat}(M)\rangle$ values the mean of the CMP distribution starts to differ from the targeted value. Therefore, we limit the validity of this regime to $\langle N_{\rm sat}(M)\rangle<0.9$ for $0.55<\nu < 4$ where the deviation are $<5\%$.

\subsection{Regime II: Large $\lambda$}
\label{sec:large regime}
\begin{figure}
    \centering
    \includegraphics[width=1\linewidth]{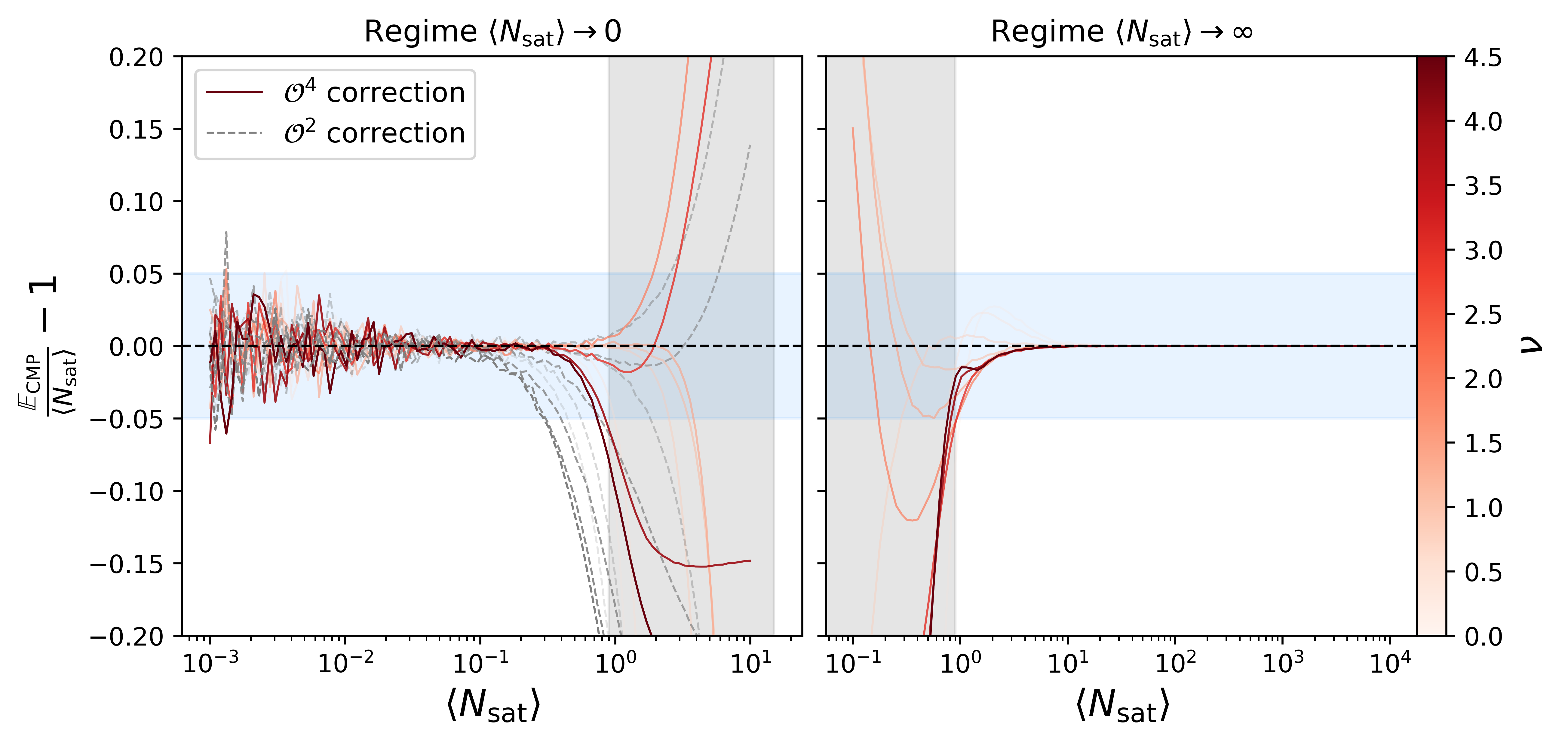}
    \caption{Comparison between $\langle N_{\rm sat}(M)\rangle$ values and the mean of the CMP distribution $\mathbb{E}_\mathrm{\lambda \to 0}$ (\textit{left panel}) and $\mathbb{E}_\mathrm{\lambda \to \infty}$ (right ) over a random sampling of size $10^6$. Each red lines show different values of the parameter $\nu = [0.5, 0.6, 0.9, 1.2, 1.5, 2, 3, 4, 4.5]$ with a colour gradient from lighter to darker with increasing $\nu$. The blue shaded regions represent the agreement within 5\% between the two quantities and the grey shaded regions show the region where the approximation is not correct any more. \textit{Left panel}: The mean of the CMP distribution $\mathbb{E}_\mathrm{CMP}$ starts to differ more than 5\% for $\langle N_{\rm sat}(M)\rangle>0.9$. For comparison grey dashed lines are similar to the red lines but using approximation to the second order $\mathcal{O}(\mathbb{E}^2)$. In this case the deviation arise at $\langle N_{\rm sat}(M)\rangle \sim 0.1$. \textit{Right panel}: The mean of the CMP distribution $\mathbb{E}_\mathrm{CMP}$ starts to differ at more than 5\% at $\langle N_{\rm sat}(M)\rangle<0.9$. }
    \label{fig: mean lambda both}

\end{figure}

For $\lambda \to \infty$, the dominant contributions arise from large value of $k$. The sum of the CMP normalisation $Z(\lambda,\nu)$ is dominated by the value of $k$ that maximizes the individual term $\lambda^k/(k!)^\nu$. The distribution becomes sharply peaked around a this specific value of $k$ and all other contributions are exponentially suppressed. In this case previous mathematical works~\cite{gaunt_asymptotic_2019} have demonstrated that the mean and variance of the CMP distribution can be approximate for any values of $\nu >0$ by:
\begin{equation}
\begin{aligned}
\label{eq:mean var CMP approx large lambda}
    \mathbb{E}[X] &= \lambda^{1/\nu} - \frac{\nu-1}{2\nu} + \frac{\nu^2-1}{24\nu^2}\lambda^{-1/\nu} 
    + \mathcal{O}(\lambda^{-1/\nu}), \\
\mathrm{Var}(X) &= \frac{\lambda^{1/\nu}}{\nu} \left( 1 - \frac{\nu^2-1}{12\nu^2}\lambda^{-1/\nu} \right) + \mathcal{O}(\lambda^{-2/\nu}).
\end{aligned}
\end{equation}
Thus at leading order it yields:
\begin{equation}
\mathbb{E}[X] \sim \lambda^{1/\nu},
\qquad
\mathrm{Var}(X) \sim \frac{\lambda^{1/\nu}}{\nu}.
\end{equation}
Unlike the Poisson distribution, where $\mathbb{E}[X]=\lambda$, in the regime $\lambda \to \infty$ the mean of the CMP distribution scales as $\lambda^{1/\nu}$.
Even if this approximation remain valid for any $\nu >0$, its accuracy depends of the couple values $(\lambda, \nu)$. Given a target mean $\mathbb{E}[X] = \langle N_{\rm sat}(M)\rangle$, one can invert Eq.~\eqref{eq:mean var CMP approx large lambda} to estimate the corresponding CMP parameter $\lambda$:
\begin{equation}
\label{eq:mean to lambda large}
    \lambda_\mathrm{\to \infty} = \left(\mathbb{E}[X] + \frac{(\nu - 1)}{2\nu}\right)^\nu
\end{equation}
Similarly to the small $\lambda$ regime we test this approximation for $\langle N_{\rm sat}(M)\rangle$ values from 0.1 to 10000. Results of the test are shown in the right panel of \Cref{fig: mean lambda both}. This approximation remains accurate at $\sim 5\%$ up to $\langle N_{\rm sat}(M)\rangle > 0.9$ for the same $\nu$ range used in the small scale approximation, $0.55<\nu < 4$. Using tabulated values we checked that taking $(\langle N_{\rm sat}(M)\rangle=0.9,\ \nu=0.5)$ leads to a 7.4\% deviation while considering $(\langle N_{\rm sat}(M)\rangle=0.9,\ \nu=0.55)$ reduced it to 4.6\%, and we can push the validity to $(\langle N_{\rm sat}(M)\rangle=0.9,\ \nu=8)$ (leading to 4.9\% deviation). Going up to $(\langle N_{\rm sat}(M)\rangle=0.9,\ \nu=8.5)$ increase the deviation to 5.6\%. Therefore, we set the limit for this approximation at $\langle N_{\rm sat}(M)\rangle > 0.9$ for $0.55<\nu < 8$. These findings are similar to what was found in~\cite{gaunt_asymptotic_2019}.

\subsection{Global scaling behaviour}
\label{sec: global behaviour}
\begin{figure}
    \centering
    \includegraphics[width=1\linewidth]{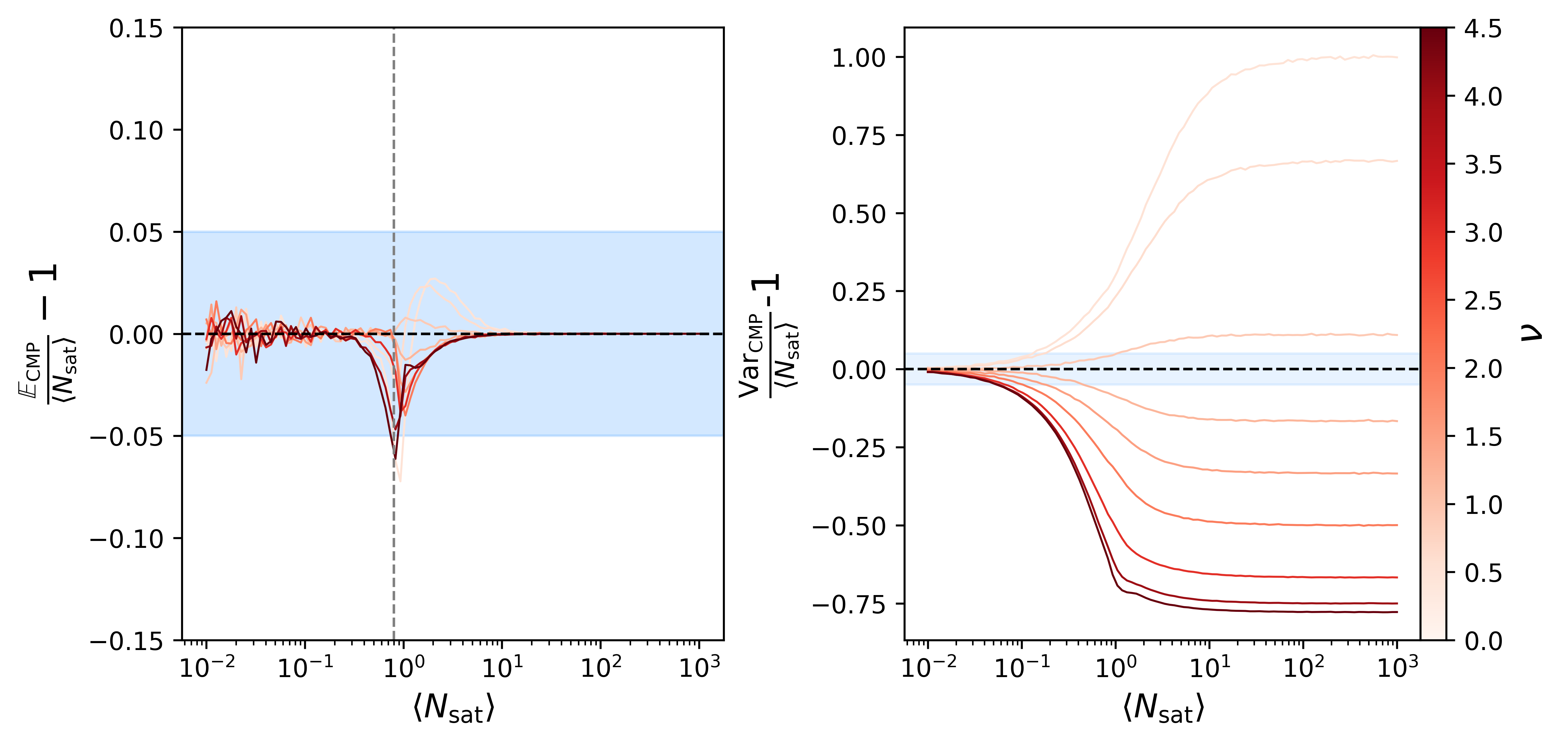}
    \caption{Comparison between $\langle N_{\rm sat}(M)\rangle$ values and the mean (left panel) and the variance (right panel) of the CMP distribution over a random sampling of size $10^6$ for the combined asymptotic behaviours. Each red lines show different values of the parameter $\nu = [0.5, 0.6, 0.9, 1.2, 1.5, 2, 3, 4, 4.5]$ with a colour gradient from lighter to darker with increasing $\nu$. The blue shaded regions represent the agreement within 5\% between the two quantities. \textit{Left panel:} The mean of the CMP distribution $\mathbb{E}$ almost agrees with the input $\langle N_{\rm sat}(M)\rangle$ value at within the 5\% region across all scales. \textit{Right panel:} The variance of the CMP distribution follows a Poisson behaviour at small $\langle N_{\rm sat}(M)\rangle$ values and starts to deviation to Poisson at $\langle N_{\rm sat}(M)\rangle~0.1$. $\nu<1$ behaves as super-poission and $\nu>1$ sub-poisson.}
    \label{fig: mean var lambda all}
\end{figure}

The asymptotic limits of the CMP distribution gives:
\begin{equation}
\mathbb{E}[X]
\sim
\begin{cases}
\lambda, & \lambda \to 0, \\
\lambda^{1/\nu}, & \lambda \to \infty,
\end{cases}
\end{equation}
and
\begin{equation}
\frac{\mathrm{Var}(X)}{\mathbb{E}[X]}
\to
\begin{cases}
\label{eq:asymptotique large lambda}
1, & \lambda \to 0, \\
1/\nu, & \lambda \to \infty.
\end{cases}
\end{equation}
The transition occurs at $\mathbb{E}[X] = \langle N_{\rm sat}(M)\rangle \sim 0.9$. In the following we construct a continuous transition between the small and large -$\lambda$ regimes. To smoothly switch between these two regimes we construct a transition based on an error function:
\begin{equation}
\begin{aligned}
\label{eq:smooth transition}
\lambda &= (1-w)\lambda_\mathrm{\to 0} + w\,\lambda_\mathrm{\to \infty}, \\
\mathrm{with} \quad w &= 0.5 \left(\frac{\langle N_{\rm sat}(M)\rangle -0.9}{0.1}-1\right)
\end{aligned}
\end{equation}
where $\lambda_\mathrm{\to 0}$ and $\lambda_\mathrm{\to \infty}$ are the expectation parameter for the two regimes from \Cref{eq:mean to lambda small} and \Cref{eq:mean to lambda large}. The transition is centre at the limit of validity of the two regimes with a smoothing scale of 0.1. The result of this continuous function for any $\langle N_{\rm sat}(M)\rangle$ values is shown in \Cref{fig: mean var lambda all}. 
The accuracy of this mapping is tested across different values of $(\langle N_{\rm sat}(M)\rangle,\nu)$ and are reported in \cref{tab:mean_mapping_accuracy}. This transition allows to recover the expected mean value from the CMP distribution on all scales within $3\%$ for $0.6<\nu$, and the $5\%$ accuracy is reached for $\nu=0.55$ at $\langle N_{\rm sat}(M)\rangle=0.9$. Going down to $\nu=0.5$ leads an underestimation of maximum $7.8\%$ at $\langle N_{\rm sat}(M)\rangle=0.9$. On the higher side the deviation reach more than $5\%$ at $\nu=5$. Thus, we impose a range limit of $0.55<\nu<4$ to stay in the $5\%$ accuracy region for any $\langle N_{\rm sat}(M)\rangle$. The right panel of \Cref{fig: mean var lambda all} displays the variance of the CMP distribution compare to its mean, showcasing the super-Poissonian behaviour at $\nu < 1$ and sub-Poissonian behaviour at $\nu > 1$ for $\lambda > 0.1$. The deviation from Poissonity approaches a constant at high occupation numbers. This means that above a given $\langle N_{\rm sat}(M)\rangle$ value, linked to a given halo mass, the variance of the CMP distribution does not change. Thus, above this given halo mass (assuming that $\langle N_{\rm sat}(M)\rangle$ increase with halo mass as it is a power law) the variance of the satellite pair counts in the 1-halo term will remain constant and will not introduce a mass-dependent divergence at large halo mass, leading to a unphysical number of satellites in a halo or an infinite amplitude of the 1 halo term.
The CMP model provides a minimal and controlled mechanism to quantify such effects through a single parameter $\nu$. Thus, fitting $\nu$ in addition to the standard HOD parameters that control the mean occupation $\langle N_{\rm sat}(M)\rangle$ provides a quantitative and minimal test of the Poisson hypothesis within the HOD framework. 

We now compare the advantages and limitations of the CMP distribution with those of the model introduced in~\cite{vos-gines_improving_2024}. In that framework they combined extended-binomial, Poisson, and negative-binomial distributions, with the nearest-integer distribution to allows for a continuous exploration of sub- and super-Poisson satellite occupations using a single dispersion parameter, $\omega$. The variance is parametrized as $\operatorname{Var}(N_{\rm sat})/\langle N_{\rm sat}(M)\rangle = 1+\omega_\sigma\langle N_{\rm sat}(M)\rangle$, where $\omega_\sigma$ denotes the effective dispersion parameter after any correction applied to avoid negative variance to the input parameter $\omega$ (see Figure 3 of~\cite{vos-gines_improving_2024}). For the CMP distribution at fixed $\nu$, the variance-to-mean ratio instead approaches $1/\nu$ at large occupation numbers. Thus, fixing the dispersion parameter in each prescription generally implies different occupation-dependent variances. Even when the two distributions have the same mean and variance, their higher-order moments can differ. One advantage of the CMP distribution is that a single distribution family covers both sub- and super-Poisson regimes, with non-negative probabilities by construction. By comparison, the framework of~\cite{vos-gines_improving_2024} combines extended-binomial, Poisson, and negative-binomial prescriptions, with the nearest-integer distribution describing the minimum-variance limit. However, the CMP distribution requires numerical evaluation of its normalization, and its mean is not directly given by $\lambda$. Matching a given HOD therefore requires an additional mapping, introducing numerical complexity and errors. The approximate mapping used in this work does not preserve the target mean occupation exactly. We impose a range limit $0.55<\nu<4$ to maintain its validity to few-percent level, up to 5\% near the boundaries, and results on mocks only lead to deviation $<1\%$ on the number of satellites galaxies relative to the Poisson case (see \Cref{sec:Impact on mocks and occupation numbers}).
\\
This model has been turn into a python package \href{https://github.com/antoine-rocher/cmpdist}{\faGithub\ \texttt{cmpdist}}, using the help of Claude Code with Opus5.5 model\footnote{The initial was a single python file with $\sim~800$ lines, all the tests and modification have been made by Claude and the outputs was verified by the author.}

\begin{table}[htbp]
    \centering
    \resizebox{\linewidth}{!}{%
    \begin{tabular}{l|cccccccc}
        \toprule
        $\langle N_{\rm sat}(M)\rangle$ & $\nu=0.4$ & $\nu=0.5$ & $\nu=0.6$ & $\nu=0.8$
                  & $\nu=1.2$ & $\nu=2$ & $\nu=4$ & $\nu=5$ \\
        \midrule
        0.05 & 0.9950 & 0.9923 & 0.9989 & 0.9990 & 0.9932 & 1.0043 & 1.0048 & 1.0062  \\
        0.2  & 1.0030 & 0.9994 & 0.9980 & 1.0016 & 1.0030 & 0.9979 & 1.0009 & 1.0008  \\
        0.5  & 0.9878 & 0.9936 & 0.9976 & 0.9987 & 0.9978 & 1.0020 & 0.9882 & 0.9786  \\
        0.85 & \textbf{0.8152} & \textbf{0.9321} & 0.9722 & 0.9987 & 0.9959 & 0.9900 & 0.9518 & \textbf{0.9328}  \\
        0.9  & \textbf{0.7639} & \textbf{0.9223} & 0.9761 & 1.0038 & 0.9918 & 0.9771 & 0.9537 & 0.9509  \\
        0.95 & \textbf{0.7582} & \textbf{0.9316} & 0.9856 & 1.0090 & 0.9881 & 0.9655 & 0.9640 & 0.9716 \\
        1.0  & \textbf{0.7868} & \textbf{0.9471} & 0.9966 & 1.0119 & 0.9865 & 0.9607 & 0.9726 & 0.9855  \\
        2.0  & 1.0209 & 1.0271 & 1.0212 & 1.0081 & 0.9934 & 0.9898 & 0.9910 & 0.9909  \\
        5.0  & 1.0175 & 1.0101 & 1.0053 & 1.0016 & 0.9999 & 0.9986 & 0.9986 & 0.9985  \\
        20   & 1.0000 & 0.9996 & 1.0000 & 1.0003 & 1.0001 & 1.0000 & 0.9998 & 1.0000  \\
        100  & 0.9998 & 1.0000 & 1.0000 & 1.0001 & 1.0000 & 1.0000 & 1.0000 & 1.0000 \\
        \bottomrule
    \end{tabular}%
    }
    \caption{Ratio of the obtained mean to the requested mean $\lambda$, for different values of $\nu$. Unity corresponds to agreement with the requested mean. All values exceeding 5\% deviation are in bold and the largest absolute departure from unity among the tabulated values occurs at $\lambda=0.95$ and $\nu=0.4$, corresponding to an underestimate of approximately $24.18\%$.}
    \label{tab:mean_mapping_accuracy}
\end{table}

\section{Impact of the CMP distribution on HOD galaxy mocks}
\label{sec:Impact on mocks}
\subsection{Impact on mocks and occupation numbers}
\label{sec:Impact on mocks and occupation numbers}

In this section we implement the CMP distribution into the HOD framework \href{https://github.com/antoine-rocher/cmpdist}{\faGithub\ \texttt{HODDIES}}~\cite{rocher_HODDIES} to evaluate the impact of the CMP on the galaxy catalogue generation and galaxy clustering. 
To perform these tests we rely on the \absum N-body simulations~\citep{maksimova_abacussummit_2021} to generate the mock catalogues. Specifically we use the snapshot at redshift z=1.1 from the \texttt{base} simulation box of size 2 Gpc/$h$ with 6912$^3$ particles. This simulation is based on a Planck 2018 $\Lambda$CDM cosmology~\cite{collaboration_planck_2020} with $h=0.6736, A_s=2.0830 \times 10^{-9}, n_s=0.9649, \omega_{cdm}=0.12, \omega_{b}=0.02237$ 
and $\sigma_8=0.8079$. 

The HOD model used for this test is the standard HOD (SHOD) --the most common HOD functional-- that uses a step function to describe central occupation and a power law to describe the satellite mean number~\cite{zheng_theoretical_2005}. This model describe galaxy samples selected by luminosity or stellar mass, where $\langle N_{\rm c}(M_h)\rangle$ is described by a smooth step function eventually reaching 1 for large-mass halo~\cite{zehavi_galaxy_2011, contreras_how_2013, smith_completed_2020, yuan_desi_2023}:
\begin{equation}
\begin{aligned}
    \left\langle N_c\left(M_h\right)\right\rangle &= \frac{A_c}{2}\left[1+\operatorname{erf}\left(\frac{\log_{10}(M_h)-\log_{10}(M_{c})}{\sigma}\right)\right], \\
    \langle N_{\rm sat}(M_h)\rangle &= A_s \left[ \frac{M_h-M_0}{M_1} \right]^{\alpha} \cdot \left\langle N_c\left(M_h\right)\right\rangle,
\label{eq:HOD LRG}
\end{aligned}
\end{equation}
where the parameters $M_{c}$ characterize the minimum halo mass to host a central galaxy, $\sigma$ describes the steepness of the transition from 0 to 1 in the number of central galaxies and $A_c$ modulates the amplitude of the distribution accounting for incomplete galaxy population. $\langle N_{\rm c}(M_h)\rangle$ must be at maximum one as only one galaxy can be at the centre of one halo, which set the upper bound of $A_c$ to one.  
$A_s$ defines the amplitude of the satellite distribution, $M_0$ is a cut-off halo mass below which no satellite galaxy can be present, $\alpha$ is the power law index that controls the variation in satellite richness with increasing halo mass, and $M_1$ is the mass at which 1 satellite is expected per halo if $A_s = 1$ and $M_0 \ll M_1$. 
For testing purposes, we fixed all HOD parameters and only vary the parameter $\nu$ of the CMP distribution to mimic a luminous red galaxy (LRG) sample of density $7\cdot10^{-4}$ [Mpc/$h]^3$ with a fraction of satellites of $\sim 15\%$. The HOD parameter are: $A_c=0.88$, $M_{c}=12.75$, $\sigma=0.5$, $A_s=0.88$, $M_0=13$, $M_1=13.5$, $\alpha=1$. All the masses are express in $\log_{10}(M_\odot/h)$. At the end, 11 mocks are created with increasing $\nu = [0.5, 0.6, 0.7, 0.8, 0.9, 1, 1.2, 1.5, 2, 3, 4]$.

\begin{figure}[ht!]
    \centering
    \includegraphics[width=\linewidth]{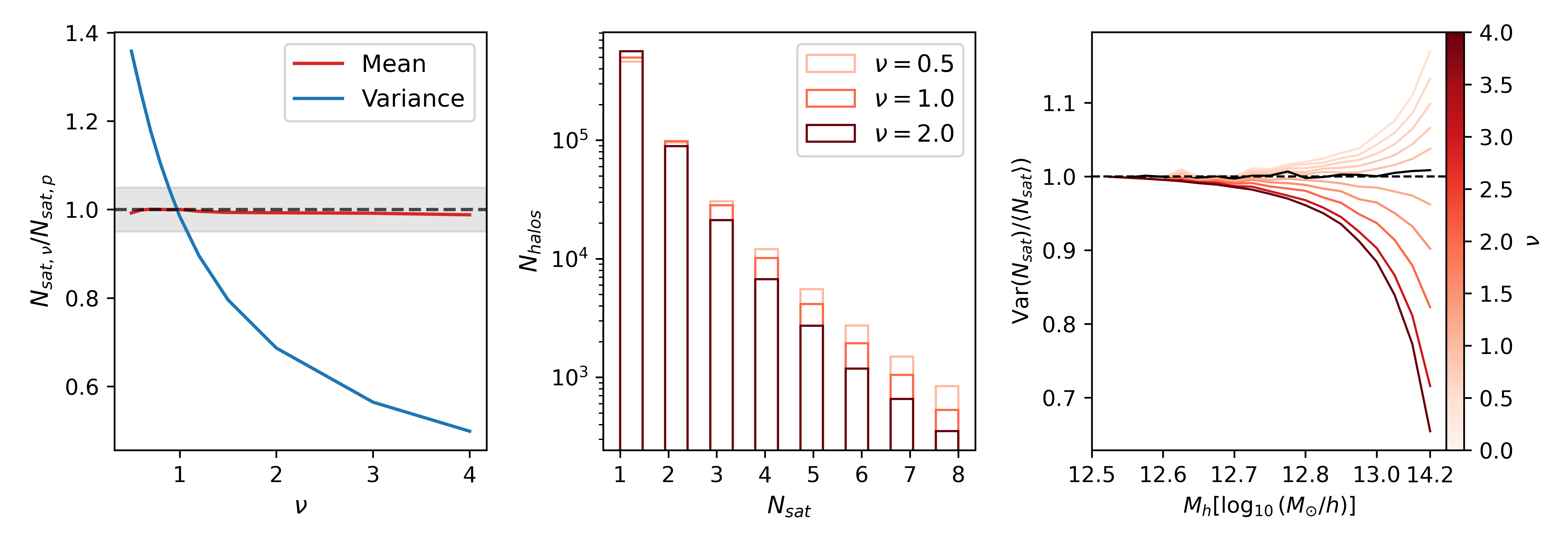}
    \caption{\textit{Left panel:} Ratio of the mean (red line) number of satellites per halo hosting at least one satellite galaxy and its variance (blue line) in the mocks, relative to the Poisson case $\nu=1$, shown as a function of the CMP dispersion parameter $\nu$. The grey band represents a $\pm 2\%$ variation with respect to the Poissonian case (dashed black line). The mean number of satellites remains stable across the $\nu$ range consider. The variance of the satellite distribution behaves as expected, decreasing as $\nu$ increases. \textit{Middle panel:} Number of halos as a function of the number of satellite galaxies per halos for three cases: $\nu=0.5$, $\nu=1$ (Poisson), and $\nu=2$. The CMP distribution produces broader satellite count distributions for $\nu<1$ and narrower distributions for $\nu>1$ compared to the Poisson case. \textit{Right panel:} Ratio of the variance and the mean of the number of satellite galaxies in bins of halo masses. Bins are divided in 10 quantiles to have the same number of satellite galaxies per halo mass bin. The Poisson case, Var($N_\mathrm{sat}$)/$\langle N_\mathrm{sat}\rangle =1$, is represented in black and other lines show the mocks with different $\nu$ values with the colour gradient from lighter to darker with increasing $\nu$.}  
    \label{fig:mock cmp distrib}
\end{figure}

The first sanity check is to validate the scaling with $\nu$ described in \Cref{sec: global behaviour}. The left panel of \Cref{fig:mock cmp distrib} shows the mean number of satellite galaxies per halo (hosting at least one satellite) in red, relative to the Poisson case. The blue curve in the left panel shows the variance of the satellite counts as a function of $\nu$, exhibiting the expected behaviour: the variance increases for $\nu<1$ (super-Poisson regime) and decreases for $\nu>1$ (sub-Poisson regime). 
The variation in the mean number of satellites remains below $\pm 1\%$ (the grey region representing a 2\% variation), which remains acceptable within the context of HOD modelling. This variation can be understood from the fact that most halos satisfy $\langle N_{\rm sat}(M_h)\rangle < 1$, placing them in the regime where the CMP distribution is highly asymmetric (i.e.\ $\lambda \to 0$). In this regime, a larger variance (corresponding to lower $\nu$) increases the probability of drawing higher satellite counts (e.g.\ $N_{\rm sat}=2,3,4$), which in turn raises the total number of satellites in the mock catalogue. Conversely, higher values of $\nu$ suppress these high-occupancy tails, leading to a smaller satellite count. The middle panel of \Cref{fig:mock cmp distrib} presents the distribution of the number of halos hosting $N_{\rm sat}$ satellite galaxies for three representative values of $\nu = [0.5,\,1,\,2]$. This provides a complementary visualization of the same effect, showing broader satellite count distributions for lower $\nu$ and narrower distributions for higher $\nu$.
Finally, we compare the mean number of satellite per halos and variance of the satellite occupation as a function of halo mass. We consider the ratio between the variance and the mean number of satellites, which directly quantifies deviations from the Poisson expectation. Satellite galaxies are grouped into halo mass bins defined using 10 quantiles, ensuring an equal number of satellites per bin. As expected, the variance is higher than the mean for $\nu < 1$ and lower for $\nu > 1$ relative to the Poisson case (shown in black), with deviations increasing toward higher halo masses. 
This behaviour arises because the deviation of the CMP distribution from the Poisson case becomes more pronounced at large $\lambda$ (see \Cref{fig: mean var lambda all}), while $\langle N_{\rm sat} \rangle$ increases with halo mass. Such trends have also been observed in hydrodynamical simulations. For example, \cite{hadzhiyska_millenniumtng_2023} (see their Figure~2) use the ratio of variance to mean to highlight departures from Poisson statistics for both star-forming and luminous red galaxy samples, particularly at high halo masses. In this context, our results demonstrate that the CMP parametrisation provides a simple and flexible framework to capture and quantify such deviations in galaxy samples.

\subsection{Implications on clustering statistics}
\label{sec:Impact on clustering}
\subsubsection{2-point correlation function}

Once the galaxy catalogues are created, we compute the clustering using the two-point correlation 2 function (2PCF). We first define the galaxy two-point correlation function in two dimensions, $\xi(r_p,\pi)$, where $\pi$ and $r_p$ are the galaxy pair separation components along and perpendicular to the line-of-sight, respectively. We then introduce the projected correlation function, $w_p(r_p)$, obtained by integrating $\xi(r_p,\pi)$ over the line-of-sight, as well as the monopole 
and quadrupole 
of the two point correlation function $\xi(s,\mu)$, where $s$ is the galaxy pair separation and $\mu$ the cosine of the angle between the line-of-sight and galaxy separation vector:  
\begin{equation}
\label{eq:clustering}
\begin{split}
w_p(r_p) &=  \int_{\pi_{min}}^{\pi_{max}} \xi(r_p,\pi) d\pi \\
\xi_l(s) &= \frac{2l+1}{2}\int_{-1}^1 \xi(s,\mu) {\it P}_l(\mu) d\mu
\end{split}
\end{equation}
where $l=[0,2]$ and ${\it P}_l(\mu)$ is the Legendre polynomial of order $l$. We rely on the DESI implementation of the Corrfunc~\cite{sinha_corrfunc_2020} package, \textsc{pycorr}\footnote{\url{https://github.com/cosmodesi/pycorr}}, to compute $\xi(r_p,\pi)$ and $\xi(s,\mu)$. For mock catalogues which are obtained from cubic boxes, the 2PCF is computed using the natural estimator~\cite{noauthor_peebles_nodate} $DD/RR-1$ which compares galaxy pair counts to the expected pair count for a uniform distribution in the box volume.
For our test we use 25 logarithmic bins in $r_p$ between 0.01 and 30 Mpc$/h$ and 80 linear bins in $\pi$ between -40 and 40 Mpc$/h$ to compute $\xi(r_p,\pi)$,. The same binning and range are used for $w_p(r_p)$ so that $\pi_{max} = - \pi_{min} = 40\,{\rm Mpc}/h$ in~\Cref{eq:clustering}. For the 2PCF multipoles, we use 25 logarithmic bins in $s$ between 0.1 and 30 Mpc$/h$ and 101 linear bins in $\mu$ between -1 and 1. 

\begin{figure}
    \label{fig: wp+xi}
    \centering
    \includegraphics[width=1\linewidth]{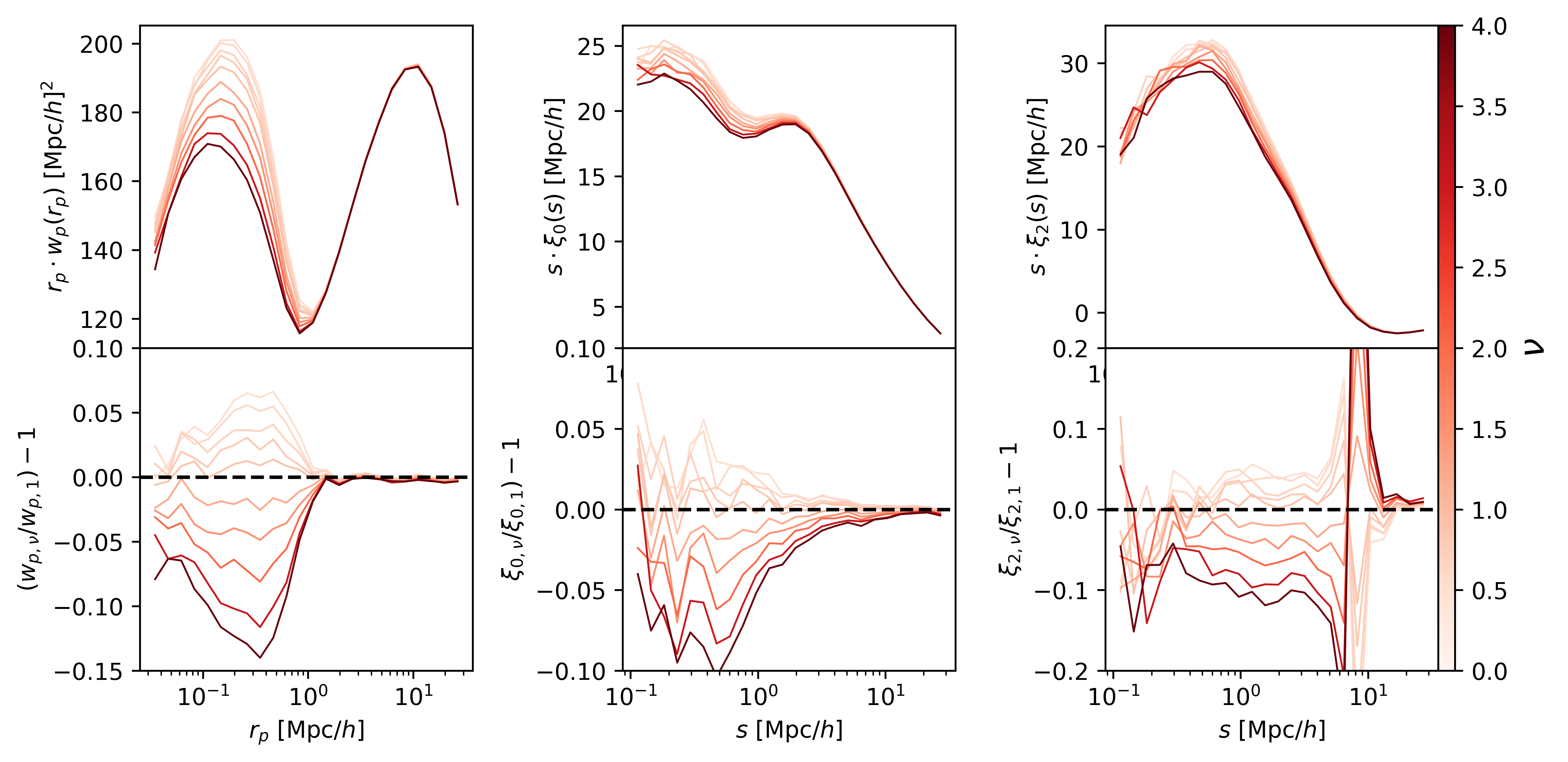}
    \caption{{\it Top panels:} Clustering measurements of the projected correlation function $w_p$ (left), the monopole (middle) and quadrupole (right) from the mocks with different values of $\nu$ with a colour gradient from lighter to darker with increasing $\nu$. {\it Bottom panels:} Ratio of the clustering results relative to the Poisson case $\nu = 1$. Small scales (1 -halo term) are impacted by the CMP dispersion parameter $\nu$ showing an increase(decrease) of clustering power for larger(smaller) variance (lower(higher) $\nu$).}
\end{figure}

The correlation function of the different mocks are displayed in \Cref{fig: wp+xi}. The upper panels shows the clustering measurements of the projected correlation function $w_p$ (left), the monopole $\xi_0$(middle) and quadrupole $\xi_2$(right) at different values of $\nu$. Changing the CMP dispersion parameter $\nu$ only impact the one-halo term at small scales as the mean number of satellites is conserved, and therefore the linear bias of the galaxy sample remain the same. A larger variance increases the probability of high-occupancy satellite realizations, thereby enhancing the one-halo contribution, as shown in \Cref{fig:mock cmp distrib}. The impact of the PDF variance has a stronger impact on the projected clustering, compare to the monopole and quadrupole. Similar behaviour has been found in former studies of the satellite PDF variance~\cite{avila_completed_2020, vos-gines_improving_2024}. However a smaller but similar trend is still present in the monopole and quadrupole with an increase of clustering power when variance is larger (smaller $\nu$) compare to the Poisson case and viceversa. This trend is expected to be significant as current spectroscopic survey (like DESI or 4MOST) can perform high precision small scale clustering measurement~\cite{rocher_desi_2023, yuan_desi_2023}.

In addition we investigate the impact of $\nu$ on the power spectrum of the 2PCF. The power spectrum of mock catalogues is evaluated with the Yamamoto estimator~\cite{yamamoto_measurement_2006} based on the Feldman-Kaiser-Peacock (FKP;~\cite{feldman_power-spectrum_1994}) field:
\begin{equation}
F(\mathbf{r})=\frac{w(\mathbf{r})\left[n_g(\mathbf{r})-\alpha n_s(\mathbf{r})\right]}{A_g^{1 / 2}}
\end{equation}
where $w(\mathbf{r})$ is the weight function. $n_g$ and $n_s$ are the observed number densities of galaxy and random catalogues, respectively. The factor $\alpha=N_g / N_s$ normalizes the contribution from the random, and the normalization factor $A$ is defined with the expected mean space density of galaxies $\bar{n}(\mathbf{r})$ as $A_g=\int d \mathbf{r}[\bar{n}(\mathbf{r}) w(\mathbf{r})]^2$. The estimator of power spectrum multipoles $P_{\ell}$ is given by:
\begin{equation}
P_{\ell}(k)=(2 \ell+1) \int \frac{d \Omega_k}{4 \pi}\left[\int d \mathbf{r}_1 \int d \mathbf{r}_2 F\left(\mathbf{r}_1\right) F\left(\mathbf{r}_2\right) \times e^{i \mathbf{k} \cdot\left(\mathbf{r}_1-\mathbf{r}_2\right)} \mathcal{P}_{\ell}\left(\hat{\mathbf{k}} \cdot \hat{\mathbf{r}}_h\right)-\mathrm{SN}_{\ell}\right]
\end{equation}
$\mathcal{P}_{\ell}$ is the Legendre polynomial and $\mathbf{r}_h=\left(\mathbf{r}_1+\mathbf{r}_2\right) / 2$. $\mathrm{SN}_{\ell}$ is the shot noise contribution expressed as $\operatorname{SN}_{\ell}(\mathbf{k})=(1+\alpha) \int d \mathbf{r} \bar{n}(\mathbf{r}) w^2(\mathbf{r}) \mathcal{L}_{\ell}(\hat{\mathbf{k}} \cdot \hat{\mathbf{r}})$, which only contributes to monopole $(\ell=0)$. In practice, we use \textsc{jaxpower}\footnote{\url{https://github.com/adematti/jax-power/tree/main/jaxpower}} to compute the power spectrum multipoles from the mock catalogues.
\begin{figure}
    \centering
    \includegraphics[width=\linewidth]{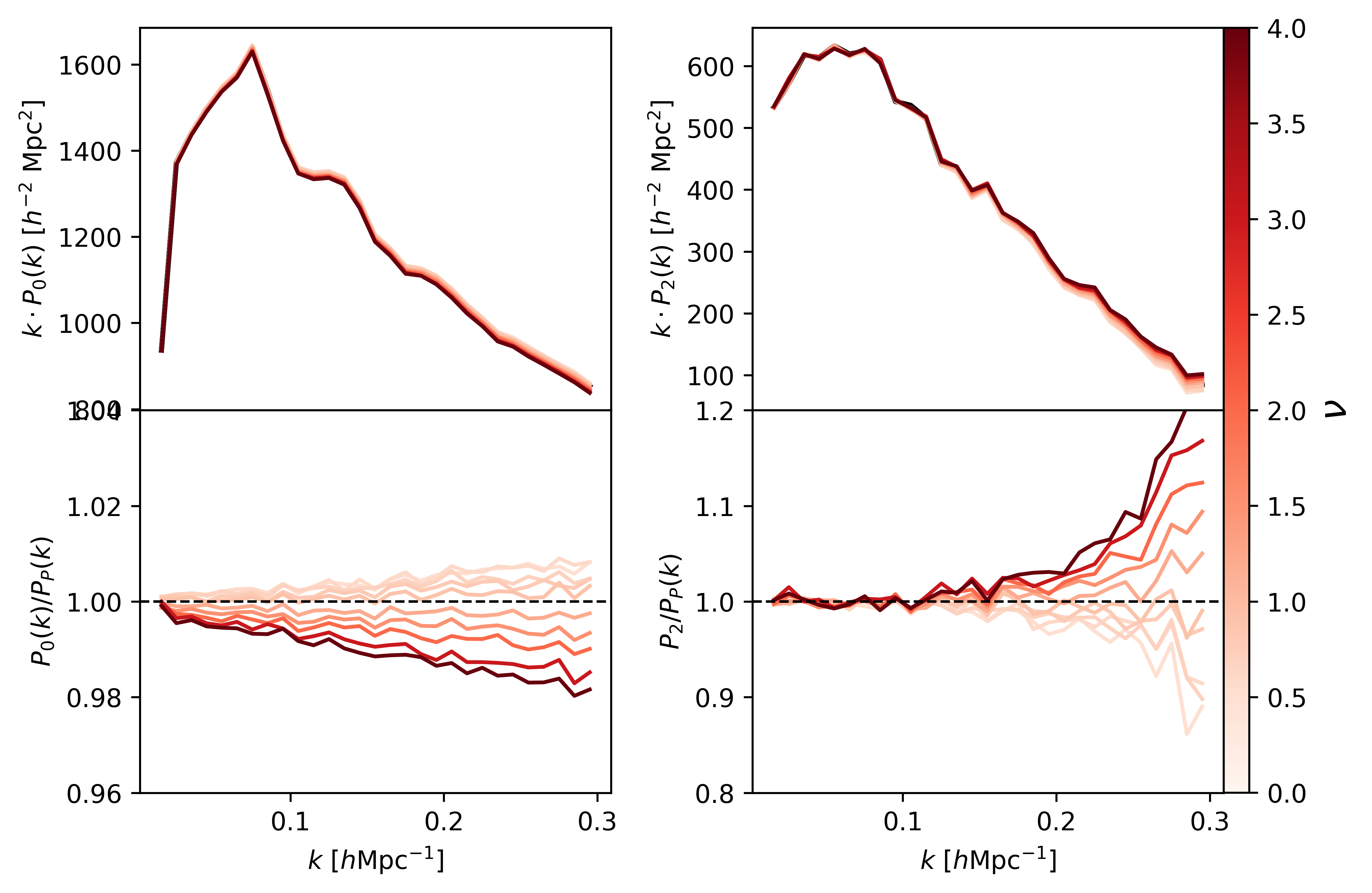}
    \caption{{\it Top panels:} Power spectrum measurements of the the monopole $P_{0}(k)$ (left) and quadrupole $P_{2}(k)$ (right) from the mocks with different values of $\nu$. The colour gradient indicate the value of increasing $\nu$ from lighter to darker. {\it Bottom panels:} Ratio of the clustering results relative to the Poisson case $\nu = 1$. }
    \label{fig:powerspectra}
\end{figure}
The results are presented in~\cref{fig:powerspectra}. The trends are similar to those found in configuration space, although the deviations are smaller. This is consistent with the larger scales probed here, with $k_\mathrm{max}=0.3\,h\,\mathrm{Mpc}^{-1}$ corresponding to a wavelength of $2\pi/k_\mathrm{max}\simeq21\,h^{-1}\mathrm{Mpc}$. We find deviations of up to $\sim2\%$ in the power spectrum monopole and $\sim10\%$ in the quadrupole. These results are broadly consistent with~\cite{avila_completed_2020}, who reported changes in the power spectrum multipoles of a few percent at $k=0.3\,h\,\mathrm{Mpc}^{-1}$, reaching $\sim10\%$ at $k=0.6\,h\,\mathrm{Mpc}^{-1}$ (see their Figure A2). Cosmological analyses commonly restrict power spectrum measurements to $k_\mathrm{max}=0.2\,h\,\mathrm{Mpc}^{-1}$~\cite{adame_desi_2025}, where our results suggest a weaker sensitivity to the satellite occupation variance. However, establishing whether these changes produce negligible cosmological parameter biases requires a dedicated inference analysis. This assessment will be particularly relevant for future analyses aiming to extend the fitted range to $k_\mathrm{max}=0.3\,h\,\mathrm{Mpc}^{-1}$~\cite{collaboration_euclid_2026} where the quadrupole can vary up to $\sim15\%$.

The above findings are expected form the halo model~\cite{asgari_halo_2023, seljak_analytic_2000}, as the 1-halo term depends explicitly on the first and second moments of the satellite distribution: 
\begin{equation}
P_{gg}^{1h}(k)
=
\frac{1}{\bar n_g^2}
\int dM \, n(M)
\Big[
2 \langle N_c N_{\rm sat} \rangle u(k|M)
+
\langle N_{\rm sat}(N_{\rm sat}-1) \rangle
|u(k|M)|^2
\Big],
\end{equation}
where $n(M)$ describe the halo mass function, $\bar n_g$ is the galaxy density, $u(k|M)$ represent the normalized Fourier transform of the halo profile. The quantity governing satellite clustering is the second factorial moment $\langle N_{\rm sat}(N_{\rm sat}-1) \rangle$ with is the Poisson assumption is equal to 
$\langle N_{\rm sat} \rangle^2$. For the CMP distribution, the small-$\lambda$ regime behave as the Poisson distribution but in the large-$\lambda$ regime, the asymptotic behaviour is $\mathrm{Var}(N_{\rm sat})\approx \frac{\langle N_{\rm sat} \rangle}{\nu}$ (see \Cref{sec: global behaviour}). For a any count distribution, $\langle N_{\rm sat}(N_{\rm sat}-1) \rangle$ can be written as a function of the variance as: 
\begin{equation}
\langle N(N-1) \rangle
=
\mathrm{Var}(N)
+
\langle N \rangle^2
-
\langle N \rangle.
\end{equation}
Using the asymptotic regime for large-$\lambda$ (\Cref{eq:asymptotique large lambda}) gives:
\begin{equation}
\langle N_{\rm sat}(N_{\rm sat}-1) \rangle
\approx
\langle N_{\rm sat} \rangle^2
+
\langle N_{\rm sat} \rangle
\left(\frac{1}{\nu} - 1\right).
\end{equation}
leading to a modification of the $\langle N_{\rm sat}(N_{\rm sat}-1) \rangle$ compare to the Poisson case that depend on $\nu$, modifying the number of satellite--satellite galaxy pairs that contributes to the one-halo term in the clustering signal.

\subsubsection{Galaxy bispectrum}

Current cosmological analyses now include three-point statistics in their fiducial clustering analysis (e.g. upcoming DESI DR2 and Euclid DR1 analysis). The three-point correlation function or the bispectrum in Fourier space, is powerful to extracting cosmological information on a given galaxy distribution, which complements that from the two-point statistics. The modelling of small scale galaxy physics can lead to modification of the observe three point signal. Therefore, in the following we quantify the impact of the satellite occupation variance on the galaxy bispectrum to assess how much it can change the bispectrum on $k$ range used for cosmological analysis. To compute the galaxy bispectrum we use the implementation of the Sugiyama estimator~\cite{sugiyama_complete_2019} in \textsc{jaxpower}. This estimator uses a decomposition of the bispectrum into the a tri-polar spherical harmonic~\cite{varshalovich_quantum_1988} to distinguish the anisotropic signal induced by the redshift space distortions (RSD;\cite{kaiser_clustering_1987}) or Alcock-Paczynski effects (AP;\cite{alcock_evolution_1979}) from the isotropic signal. The bispectrum $B\left(\boldsymbol{k}_1, \boldsymbol{k}_2, \hat{\boldsymbol{n}}\right)$ as function of two wavevectors $\boldsymbol{k}_1, \boldsymbol{k}_2$ and the line-of-sight vector $\hat{n}$ in the under the local plane-parallel approximation is given as:
\begin{equation}
B\left(\boldsymbol{k}_1, \boldsymbol{k}_2, \hat{\boldsymbol{n}}\right)=\sum_{\ell_1, \ell_2, L} B_{\ell_1 \ell_2 L}\left(k_1, k_2\right) S_{\ell_1 \ell_2 L}\left(\hat{\boldsymbol{k}}_1, \hat{\boldsymbol{k}}_2, \hat{\boldsymbol{n}}\right), 
\end{equation}
assuming parity symmetry condition that restricts $\ell_1+ \ell_2+ L$
to even numbers. $S_{\ell_1 \ell_2 L}\left(\hat{\boldsymbol{k}}_1, \hat{\boldsymbol{k}}_2, \hat{\boldsymbol{n}}\right)$ is the tripolar spherical harmonics basis~\cite{varshalovich_quantum_1988} define as:
\begin{equation}
S_{\ell_1 \ell_2 L}\left(\hat{\boldsymbol{k}}_1, \hat{\boldsymbol{k}}_2, \hat{\boldsymbol{n}}\right)=\frac{1}{H_{\ell_1 \ell_2 L}} \sum_{m_1 m_2 M}\left(\begin{array}{ccc}
\ell_1 & \ell_2 & L \\
m_1 & m_2 & M
\end{array}\right) y_{\ell_1}^{m_1}\left(\hat{\boldsymbol{k}}_1\right) y_{\ell_2}^{m_2}\left(\hat{\boldsymbol{k}}_2\right) y_L^M(\hat{\boldsymbol{n}}).
\end{equation}
where coefficients $H$ are defined in terms of Wigner $3-j$ symbols as $H_{\ell_1 \ell_2 L}=\left(\begin{array}{ccc}\ell_1 & \ell_2 & L \\ 0 & 0 & 0\end{array}\right)$ and $y_{\ell}^m=\sqrt{\frac{4 \pi}{2 \ell+1}} Y_{\ell}^m$ are the reduced spherical harmonics. 
The multipole coefficients of the bispectrum are given by:
\begin{equation}
\begin{aligned}
B_{\ell_1 \ell_2 L}\left(k_1, k_2\right) & =N_{\ell_1 \ell_2 L} H_{\ell_1 \ell_2 L}  \sum_{m_1 m_2 M}\left(\begin{array}{ccc}
\ell_1 & \ell_2 & L \\
m_1 & m_2 & M
\end{array}\right)\\ & \quad \times \int \frac{\mathrm{d}^2 \hat{\mathbf{k}}_1}{4 \pi} \frac{\mathrm{~d}^2 \hat{\mathbf{k}}_2}{4 \pi} \frac{\mathrm{~d}^2 \hat{\mathbf{n}}}{4 \pi} y_{\ell_1}^{m_1 *}\left(\hat{\mathbf{k}}_1\right) y_{\ell_2}^{m_2 *}\left(\hat{\mathbf{k}}_2\right) y_L^{M^*}(\hat{\mathbf{n}}) B\left(\mathbf{k}_1, \mathbf{k}_2, \hat{\mathbf{n}}\right)
\end{aligned}
\end{equation}
with the prefactor $N_{\ell_1 \ell_2 L}=\left(2\ell_1+1\right)\left(2 \ell_2+1\right)(2 L+1)$. The multipole index $L$ characterizes the anisotropy of the bispectra along the LOS direction induced by the RSD or AP effect. $B_{000}$ and $B_{202}$ are the dominant monopole ($L=0$) and quadrupole ($L=2$) moments, respectively, and are expected to yield the largest signal-to-noise ratio~\cite{sugiyama_complete_2019}. 

\begin{figure}[t]
    \centering
    \includegraphics[width=\linewidth]{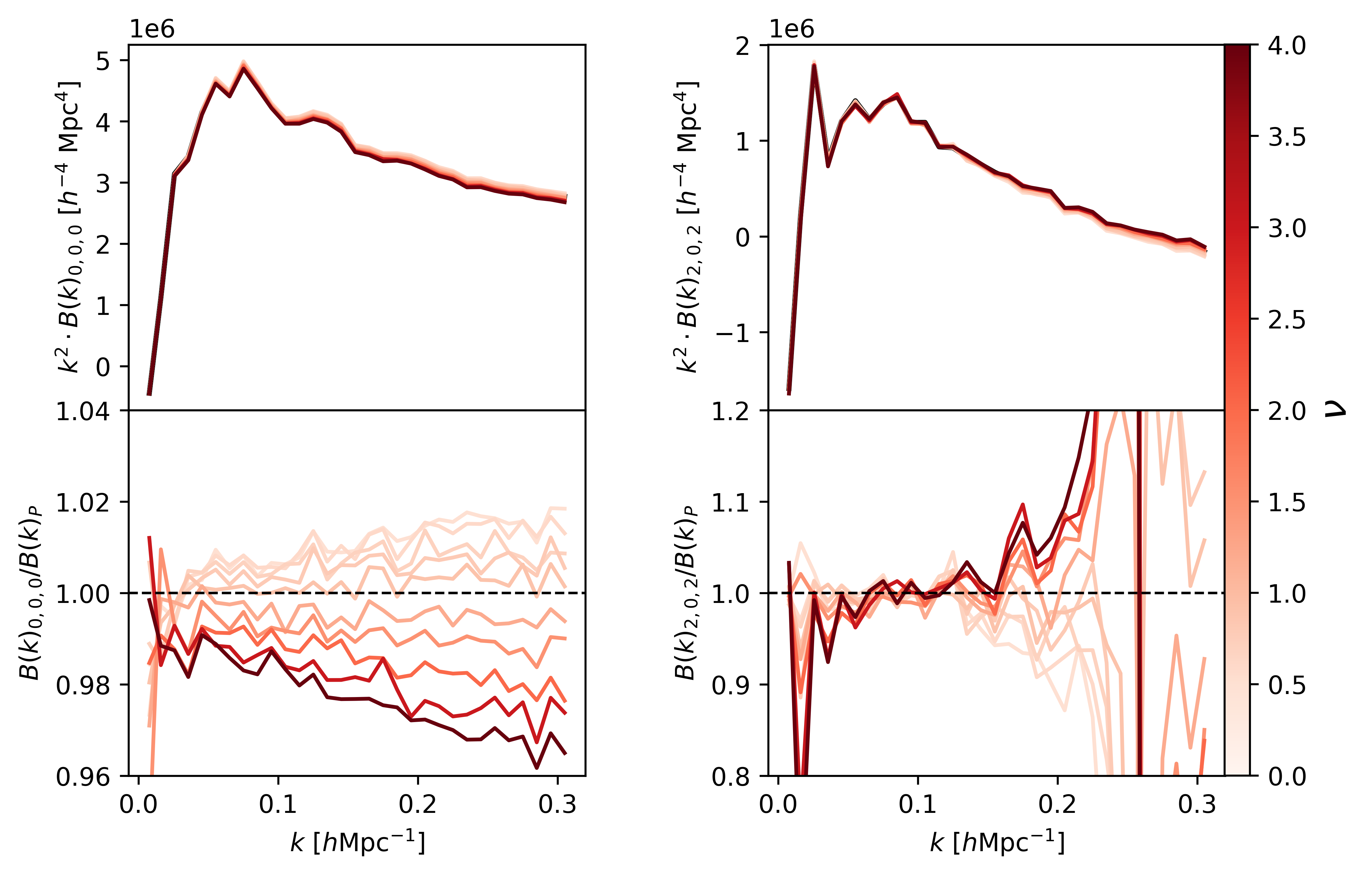}
    \caption{{\it Top panels:} Bispectrum measurements of the the monopole $B_{000}$ (left)  and quadrupole $B_{202}$ (right) from the mocks with different values of $\nu$. The colour gradient indicate the value of increasing $\nu$ from lighter to darker. {\it Bottom panels:} Ratio of the clustering results relative to the Poisson case $\nu = 1$. }
    \label{fig:bispectra}
\end{figure}

Similarly to the previous section we computed the bispectrum multipoles $B_{000}$ and $B_{202}$ for our 11 mocks with varying $\nu \in [0.5,4]$. We use 20 bins linearly space within a range of $k \in [0, 0.3]$ [$h$/Mpc] with a 400 size mesh grid. The results are shown in \Cref{fig:bispectra}. 
Interestingly, the bispectrum does not show strong variation when varying the CMP dispersion parameter $\nu$.We observe a $2\%$ variation for super-Poisson cases ($\nu<1$) and up to 4\% for sub-Poisson ($\nu>1$) cases on the monopole $B_{000}$ at $k_\mathrm{max}=0.3$. This deviation are small compare to what we can expected from data errors. For the quadrupole $B_\mathrm{202}$, measurements from mocks are nosier but the overall trend is similar to the monopole with only small variation with $\nu$ and the expected data errors are also expected to be larger. Thus, the bispectrum seems to be rather insensitive against the change of Poisson assumption for $k_\mathrm{max} < 0.3$ and therefore we do not expect strong biases on cosmological constrains. However, a detailed analysis has to be done to investigate more on the impact of non Poisson satellite distribution on cosmological constrain using the bispectrum, and would be the purpose of a future work.  

\subsubsection{Count-in-cylinders}

Count-in-cylinders (CIC;~\cite{reid_constraining_2009}) statistics are sensitive to higher-order clustering and therefore provide a powerful complement to the two-point correlation function. The CIC statistic is defined as the probability distribution of the number of companion galaxies found within cylinders of radius $R_{\rm C}$ and length $L_{\rm C}$ centred on galaxies. As such, it probes the local galaxy environment and is particularly sensitive to variations in the variance of the satellite occupation. Unlike two-point statistics, the CIC explicitly depends on the galaxy number density, since the number of companions within a fixed volume scales with the underlying density field. Recently, CIC statistics have been used to constrain HOD parameters and evidenced positive sign of assembly bias in low redshift galaxy sample from SDSS and DESI~\cite{wang_evidence_2022, pearl_desi_2023}.

\begin{figure}
    \centering
    \includegraphics[width=\linewidth]{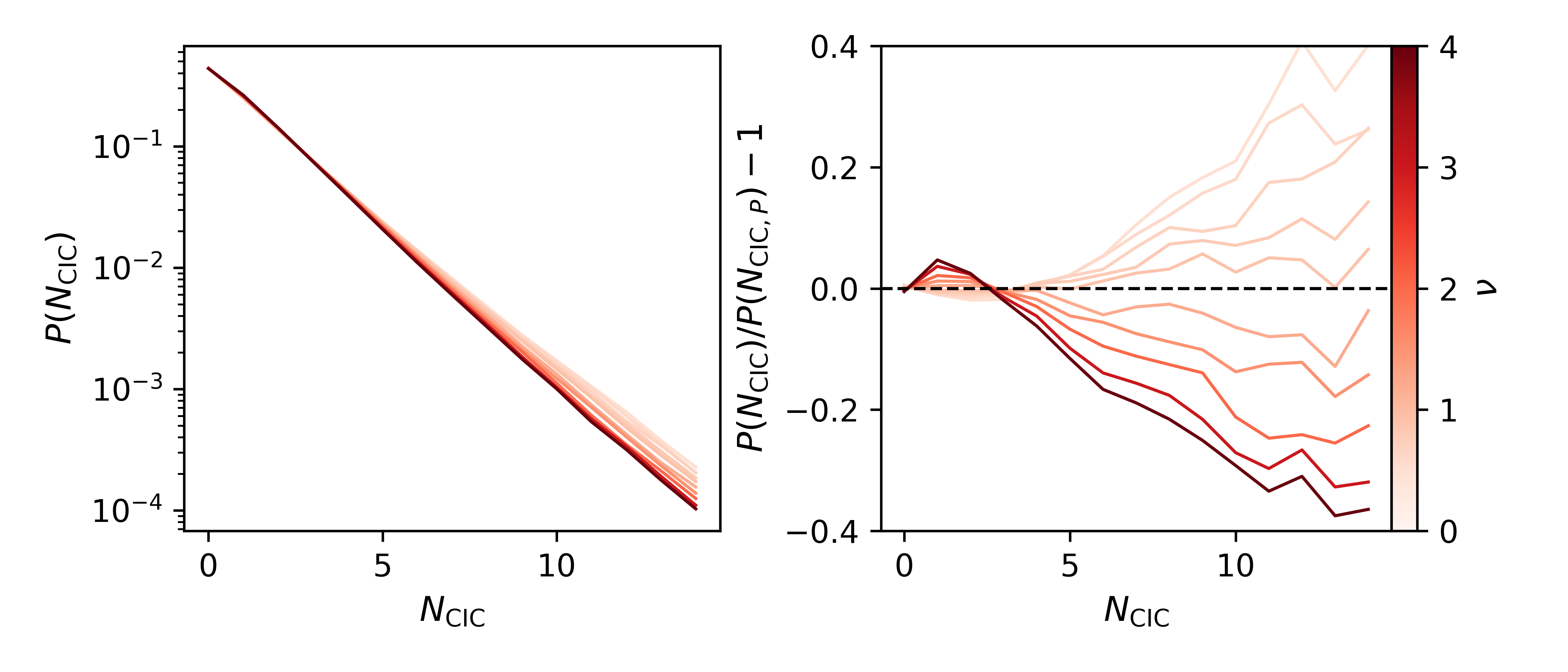}
    \caption{\textit{Left panel:} Number counts of cylinder $P(N_\mathrm{CIC})$ as a function of the number of galaxy counts enclosed in a cylinder $N_\mathrm{CIC}$ up to 15. \textit{Right panel:} Ratio between all $P(N_\mathrm{CIC})$ and the Poisson case  $P(N_\mathrm{CIC, P})$ ($\nu=1$). Different values of $\nu$ are represented with a colour gradient from lighter to darker with increasing $\nu$.}
    \label{fig:CIC}
\end{figure}

In the following we investigate the impact of the satellite occupation variance on the CIC statistic. The number of galaxy counts enclosed in a cylinder $N_\mathrm{CIC}$ is performed using cylinders of radius $R_\mathrm{C}=2$ and length $L_\mathrm{C}=10$ centred on each galaxy of our sample, similarly to previous work from~\cite{wang_evidence_2022, pearl_desi_2023}. We also exclude self-counting allowing $N_\mathrm{CIC}=0$. Then, the CIC distribution $P(N_\mathrm{CIC})$ is evaluated in linear bins of $N_\mathrm{CIC}$ between -0.5 and 15. The results are presented in \Cref{fig:CIC}. The number counts of cylinder $P(N_\mathrm{CIC})$ as a function of the number of galaxy counts enclosed in a cylinder $N_\mathrm{CIC}$ is display in the left panel for mocks with different $\nu$ values. The right panel shows the relative difference between all $P(N_\mathrm{CIC})$ and the Poisson case $P(N_\mathrm{CIC, P})$ ($\nu=1$). The CIC distribution shows a higher number of cylinder with higher galaxy counts and oppositely, a smaller number of cylinder with lower galaxy counts  when the variance of the satellite occupation is larger ($\nu<1$) compare to the Poisson case. This is expected and consistent with the results using the 2PCF. However, the impact of the satellite PDF variance is stronger for CIC statistic compare to the 2PCF, with variation up to $~40\%$ relative to the Poisson case. This highlight the sensitivity of CIC to the higher-order moments of the galaxy distribution and to test deviations from Poisson statistics in satellite populations. This highlights the sensitivity of the CIC statistic to higher-order moments of the galaxy distribution, showing its ability to probe deviations from Poisson satellite statistics. These trends agree qualitatively with the count-in-cells results in~\cite{vos-gines_improving_2024}, which showed that increasing the satellite occupation variance enhances the numbers of galaxies per cells, while decreasing the variance produces the opposite effect. They also demonstrated, through fits to a reference mock catalogue, that count-in-cells can constrain the dispersion parameter more strongly than the two-point statistics considered in their analysis. Our results support this sensitivity to non-Poisson satellite occupations using count-in-cylinders although we do not perform a corresponding parameter-inference analysis in this study and leave it for future work.



\section{Conclusion}
\label{sec:conclusion}

This work presents the Conway--Maxwell--Poisson (CMP) distribution as a minimal and flexible generalization of the Poisson distribution within the Halo Occupation Distribution (HOD) framework. The CMP distribution introduces a single additional parameter, the dispersion parameter $\nu$, which allows to control the variation between sub-Poisson ($\nu > 1$) and super-Poisson ($\nu < 1$) regimes. One limitation of the CMP distribution is the numerical estimation of the normalizing constant $Z(\lambda, \nu)$ (\Cref{eq:normalisation constant}) which has no closed form. We design a scale dependent way to numerically compute $Z(\lambda, \nu)$ that ensure to reach the convergence for any value of $\lambda$ and $0.55<\nu <4$. In contrast to the Poisson distribution, the mean of the CMP distribution $\mathbb{E}_\mathrm{CMP}$ is not equal to its expectation parameter $\lambda_\mathrm{CMP}$ (\Cref{eq: mean var CMP}), which complicates its direct use in HOD modelling where the mean satellite occupation $\langle N_{\rm sat}(M)\rangle$ must be preserved. To overcome this, we derive analytical approximations that allow us to invert the CMP mean and recover the value of $\lambda$ corresponding to a target mean occupation. We show that the CMP distribution exhibits distinct behaviours depending on the value of $\lambda$. In the small-$\lambda$ regime (\Cref{sec:small regime}), we derive a series expansion up to $\mathcal{O}(\lambda^4)$, which extends the validity of this approximation up to $\lambda \lesssim 0.9$ with $\sim 5\%$ accuracy for $0.55 < \nu < 4$. In the large-$\lambda$ regime (\Cref{sec:large regime}), we adopt asymptotic expansions from previous work~\cite{gaunt_asymptotic_2019}, which are accurate at the $\sim 5\%$ level for $\lambda \gtrsim 0.9$ over a range $0.55<\nu<8$. We then construct a smooth interpolation between these regimes (\Cref{sec: global behaviour}), resulting in a continuous and accurate mapping between $\lambda$ and $\langle N_{\rm sat}(M)\rangle$. Overall, this provides an efficient and robust method to control the variance of the satellite occupation while preserving its mean. We find that deviations from Poisson behaviour become significant for $\lambda \gtrsim 0.1$, with the variance asymptotically scaling as $\mathrm{Var}(N) \sim \langle N \rangle / \nu$ at large $\lambda$.

In the second part of this work, we implement the CMP distribution within the \texttt{HODDIES} code~\cite{rocher_HODDIES} and construct mock galaxy catalogues with to quantify the impact of non-Poisson satellite statistics on galaxy occupation and clustering observables. In \Cref{sec:halo occupation statistics}, we first validate our implementation by examining how the satellite population varies with the dispersion parameter $\nu$. We find that the mean number of satellites remains within $\pm 1\%$ as a function of $\nu$, consistent with expectations from the asymmetry of the CMP distribution at low $\lambda$. As expected, the variance of the satellite occupation increases for $\nu < 1$ (super-Poisson) and decreases for $\nu > 1$ (sub-Poisson). This behaviour leads to a broader (narrower) distribution of satellite counts in halos for $\nu < 1$ ($\nu > 1$). We also observe that these deviations become more pronounced with increasing halo mass, reflecting the stronger departure from Poisson statistics at larger $\lambda$. Similar mass-dependent deviations have been reported in hydrodynamical simulations~\cite{hadzhiyska_millenniumtng_2023}, highlighting the relevance of the CMP framework for modelling realistic galaxy populations. Variations in the satellite occupation distribution directly impact small-scale clustering. In particular, broader satellite distributions enhance the one-halo term of the correlation function, leading to an increase in clustering amplitude at scales $r_p \lesssim 1\,h^{-1}\mathrm{Mpc}$. In \Cref{sec:Impact on clustering}, we quantify these effects by measuring the monopole $\xi_0$, quadrupole $\xi_2$, and projected correlation function $w_p$. We find that deviations from Poisson statistics lead to relative differences of up to $\sim 15\%$ in $w_p$ and $\sim 10\%$ in $\xi_0$ and $\xi_2$ at small scales. While moderate, these effects can be significant compare to the precision of current observational datasets. 
We also investigate the impact on the power spectrum multipoles up to $k_\mathrm{max} = 0.3$ and found small deviation up to $\sim2\%$ in the power spectrum monopole and $\sim10\%$ for the quadrupole, comparable to the results reported in~\cite{avila_completed_2020}.
We further investigated the impact of $\nu$ on bispectrum measurements using the Sugiyama basis~\cite{sugiyama_complete_2019}. We found bispectrum measurements to be robust against the change of Poisson assumption for satellite occupation with a small trend, less than $2\%$, on the bispectrum monopole up to $k_\mathrm{max}=0.3$. These variations are expected to be smaller than one could expect from the bispectrum data errors at these scales and should not bias cosmological analysis. However a detailed analysis should be conduct to confirm this statement and would be part of a future work. Finally, we investigate counts-in-cylinders (CIC) statistics as a complementary probe of small-scale clustering and higher-order moments of the galaxy distribution. Unlike two-point statistics, CIC directly probe the local galaxy environment and are particularly sensitive to variations in the variance of the satellite occupation. We find that CIC statistics exhibit larger deviations from the Poisson case, with differences reaching up to $\sim 40\%$. This highlights their strong potential for constraining non-Poisson satellite statistics. 
These results complement those of~\cite{vos-gines_improving_2024}. The sensitivity of small-scale two-point clustering and CIC statistics to non-Poisson satellite occupations confirms the qualitative behaviour established in their work. The novelty of this work is the introduction of the CMP which provide a single distribution and a different variance–occupation relation compare to~\cite{vos-gines_improving_2024}. However, the approximate mapping presented in this work reproduces the target mean occupation with some small variations up to 1\% for $\nu=4$. We additionally investigate galaxy bispectrum multipoles, finding a comparatively weak response, at approximately the (2\%) level up to ($k=0.3,h,{\rm Mpc}^{-1}$), for the LRG-like HOD and dispersion range considered. This extends the comparison of clustering observables beyond those examined in~\cite{vos-gines_improving_2024}. Further opportunities to probe these effects may arise from recently proposed higher-order statistics, including galaxy multiplets~\cite{wang_galaxy-multiplet_2025}, which could be particularly sensitive to deviations from Poisson behaviour.
Overall, our results demonstrate that the CMP parametrisation provides a robust and efficient framework to probe deviations from Poisson satellite statistics in galaxy surveys. It offers a direct and minimally extended way to test the Poisson assumption within the HOD framework using current and upcoming datasets. As cosmological analyses reach increasing precision, incorporating such extensions will be essential to control systematic uncertainties associated with the galaxy--halo connection. Future work will focus on constraining the CMP parameter $\nu$ using observational data from DESI, exploring its dependence on galaxy properties and halo environment, and assessing the impact of non-Poisson satellite statistics on higher-order clustering observables and cosmological parameter inference.

\acknowledgments

AR thanks Rafaela Gsponer, Charles Dalang and Shengyu HE for useful discussions and constructive comment on this manuscript.
AR acknowledge support from the Swiss National Science Foundation (SNF) “Cosmology with 3D Maps of the Universe” research grant 200020\_207379.


\bibliographystyle{JHEP}
\bibliography{ref}{}

@article{berlind_halo_2002,
	title = {The {Halo} {Occupation} {Distribution}: {Towards} an {Empirical} {Determination} of the {Relation} {Between} {Galaxies} and {Mass}},
	volume = {575},
	issn = {0004-637X, 1538-4357},
	shorttitle = {The {Halo} {Occupation} {Distribution}},
	url = {http://arxiv.org/abs/astro-ph/0109001},
	doi = {10.1086/341469},
	number = {2},
	urldate = {2025-11-03},
	journal = {The Astrophysical Journal},
	author = {Berlind, Andreas A. and Weinberg, David H.},
	month = aug,
	year = {2002},
	note = {arXiv:astro-ph/0109001},
	pages = {587--616},
}

@article{collaboration_desi_2025,
	title = {{DESI} {DR2} {Results} {II}: {Measurements} of {Baryon} {Acoustic} {Oscillations} and {Cosmological} {Constraints}},
	volume = {112},
	issn = {2470-0010, 2470-0029},
	shorttitle = {{DESI} {DR2} {Results} {II}},
	url = {http://arxiv.org/abs/2503.14738},
	doi = {10.1103/tr6y-kpc6},
	language = {en},
	number = {8},
	urldate = {2025-11-03},
	journal = {Physical Review D},
	author = {Collaboration, DESI and Abdul-Karim, M. and Aguilar, J. and Ahlen, S. and Alam, S. and Allen, L. and Prieto, C. Allende and Alves, O. and Anand, A. and Andrade, U. and Armengaud, E. and Aviles, A. and Bailey, S. and Baltay, C. and Bansal, P. and Bault, A. and Behera, J. and BenZvi, S. and Bianchi, D. and Blake, C. and Brieden, S. and Brodzeller, A. and Brooks, D. and Buckley-Geer, E. and Burtin, E. and Calderon, R. and Canning, R. and Rosell, A. Carnero and Carrilho, P. and Casas, L. and Castander, F. J. and Cereskaite, R. and Charles, M. and Chaussidon, E. and Chaves-Montero, J. and Chebat, D. and Chen, X. and Claybaugh, T. and Cole, S. and Cooper, A. P. and Cuceu, A. and Dawson, K. S. and Macorra, A. de la and Mattia, A. de and Deiosso, N. and Costa, J. Della and Demina, R. and Dey, A. and Dey, B. and Ding, Z. and Doel, P. and Edelstein, J. and Eisenstein, D. J. and Elbers, W. and Fagrelius, P. and Fanning, K. and Fernández-García, E. and Ferraro, S. and Font-Ribera, A. and Forero-Romero, J. E. and Frenk, C. S. and Garcia-Quintero, C. and Garrison, L. H. and Gaztañaga, E. and Gil-Marín, H. and Gontcho, S. Gontcho A. and Gonzalez, D. and Gonzalez-Morales, A. X. and Gordon, C. and Green, D. and Gutierrez, G. and Guy, J. and Hadzhiyska, B. and Hahn, C. and He, S. and Herbold, M. and Herrera-Alcantar, H. K. and Ho, M. and Honscheid, K. and Howlett, C. and Huterer, D. and Ishak, M. and Juneau, S. and Kamble, N. V. and Karaçaylı, N. G. and Kehoe, R. and Kent, S. and Kim, A. G. and Kirkby, D. and Kisner, T. and Koposov, S. E. and Kremin, A. and Krolewski, A. and Lahav, O. and Lamman, C. and Landriau, M. and Lang, D. and Lasker, J. and Goff, J. M. Le and Guillou, L. Le and Leauthaud, A. and Levi, M. E. and Li, Q. and Li, T. S. and Lodha, K. and Lokken, M. and Lozano-Rodríguez, F. and Magneville, C. and Manera, M. and Martini, P. and Matthewson, W. L. and Meisner, A. and Mena-Fernández, J. and Menegas, A. and Mergulhão, T. and Miquel, R. and Moustakas, J. and Muñoz-Gutiérrez, A. and Muñoz-Santos, D. and Myers, A. D. and Nadathur, S. and Naidoo, K. and Napolitano, L. and Newman, J. A. and Niz, G. and Noriega, H. E. and Paillas, E. and Palanque-Delabrouille, N. and Pan, J. and Peacock, J. and Ibanez, Marcos Pellejero and Percival, W. J. and Pérez-Fernández, A. and Pérez-Ràfols, I. and Pieri, M. M. and Poppett, C. and Prada, F. and Rabinowitz, D. and Raichoor, A. and Ramírez-Pérez, C. and Rashkovetskyi, M. and Ravoux, C. and Rich, J. and Rocher, A. and Rockosi, C. and Rohlf, J. and Román-Herrera, J. O. and Ross, A. J. and Rossi, G. and Ruggeri, R. and Ruhlmann-Kleider, V. and Samushia, L. and Sanchez, E. and Sanders, N. and Schlegel, D. and Schubnell, M. and Seo, H. and Shafieloo, A. and Sharples, R. and Silber, J. and Sinigaglia, F. and Sprayberry, D. and Tan, T. and Tarlé, G. and Taylor, P. and Turner, W. and Ureña-López, L. A. and Vaisakh, R. and Valdes, F. and Valogiannis, G. and Vargas-Magaña, M. and Verde, L. and Walther, M. and Weaver, B. A. and Weinberg, D. H. and White, M. and Wolfson, M. and Yèche, C. and Yu, J. and Zaborowski, E. A. and Zarrouk, P. and Zhai, Z. and Zhang, H. and Zhao, C. and Zhao, G. B. and Zhou, R. and Zou, H.},
	month = oct,
	year = {2025},
	note = {arXiv:2503.14738 [astro-ph]},
	pages = {083515},
}

@article{findlay_exploring_2025,
	title = {Exploring {HOD}-dependent systematics for the {DESI} 2024 {Full}-{Shape} galaxy clustering analysis},
	volume = {2025},
	issn = {1475-7516},
	url = {http://arxiv.org/abs/2411.12023},
	doi = {10.1088/1475-7516/2025/09/007},
	language = {en},
	number = {09},
	urldate = {2025-11-03},
	journal = {Journal of Cosmology and Astroparticle Physics},
	author = {Findlay, N. and Nadathur, S. and Percival, W. J. and Mattia, A. de and Zarrouk, P. and Gil-Marín, H. and Alves, O. and Mena-Fernández, J. and Garcia-Quintero, C. and Rocher, A. and Ahlen, S. and Bianchi, D. and Brooks, D. and Claybaugh, T. and Cole, S. and Macorra, A. de la and Dey, A. and Doel, P. and Fanning, K. and Font-Ribera, A. and Forero-Romero, J. E. and Gaztañaga, E. and Gutierrez, G. and Hahn, C. and Honscheid, K. and Howlett, C. and Juneau, S. and Levi, M. E. and Meisner, A. and Miquel, R. and Moustakas, J. and Palanque-Delabrouille, N. and Pérez-Ràfols, I. and Rossi, G. and Sanchez, E. and Schlegel, D. and Schubnell, M. and Seo, H. and Sprayberry, D. and Tarlé, G. and Vargas-Magaña, M. and Weaver, B. A.},
	month = sep,
	year = {2025},
	note = {arXiv:2411.12023 [astro-ph]},
	pages = {007},
}

@article{adame_desi_2025,
	title = {{DESI} 2024 {V}: {Full}-{Shape} {Galaxy} {Clustering} from {Galaxies} and {Quasars}},
	volume = {2025},
	issn = {1475-7516},
	shorttitle = {{DESI} 2024 {V}},
	url = {http://arxiv.org/abs/2411.12021},
	doi = {10.1088/1475-7516/2025/09/008},
	language = {en},
	number = {09},
	urldate = {2025-11-03},
	journal = {Journal of Cosmology and Astroparticle Physics},
	author = {Adame, A. G. and Aguilar, J. and Ahlen, S. and Alam, S. and Alexander, D. M. and Alvarez, M. and Alves, O. and Anand, A. and Andrade, U. and Armengaud, E. and Avila, S. and Aviles, A. and Awan, H. and Bailey, S. and Baltay, C. and Bault, A. and Behera, J. and BenZvi, S. and Beutler, F. and Bianchi, D. and Blake, C. and Blum, R. and Brieden, S. and Brodzeller, A. and Brooks, D. and Buckley-Geer, E. and Burtin, E. and Calderon, R. and Canning, R. and Rosell, A. Carnero and Cereskaite, R. and Cervantes-Cota, J. L. and Chabanier, S. and Chaussidon, E. and Chaves-Montero, J. and Chen, S. and Chen, X. and Claybaugh, T. and Cole, S. and Cuceu, A. and Davis, T. M. and Dawson, K. and Macorra, A. de la and Mattia, A. de and Deiosso, N. and Dey, A. and Dey, B. and Ding, Z. and Doel, P. and Edelstein, J. and Eftekharzadeh, S. and Eisenstein, D. J. and Elliott, A. and Fagrelius, P. and Fanning, K. and Ferraro, S. and Ereza, J. and Findlay, N. and Flaugher, B. and Font-Ribera, A. and Forero-Sánchez, D. and Forero-Romero, J. E. and Garcia-Quintero, C. and Garrison, L. H. and Gaztañaga, E. and Gil-Marín, H. and Gontcho, S. Gontcho A. and Gonzalez-Morales, A. X. and Gonzalez-Perez, V. and Gordon, C. and Green, D. and Gruen, D. and Gsponer, R. and Gutierrez, G. and Guy, J. and Hadzhiyska, B. and Hahn, C. and Hanif, M. M. S. and Herrera-Alcantar, H. K. and Honscheid, K. and Howlett, C. and Huterer, D. and Iršič, V. and Ishak, M. and Juneau, S. and Karaçaylı, N. G. and Kehoe, R. and Kent, S. and Kirkby, D. and Kong, H. and Koposov, S. E. and Kremin, A. and Krolewski, A. and Lai, Y. and Lan, T.-W. and Landriau, M. and Lang, D. and Lasker, J. and Goff, J. M. Le and Guillou, L. Le and Leauthaud, A. and Levi, M. E. and Li, T. S. and Lodha, K. and Magneville, C. and Manera, M. and Margala, D. and Martini, P. and Maus, M. and McDonald, P. and Medina-Varela, L. and Meisner, A. and Mena-Fernández, J. and Miquel, R. and Moon, J. and Moore, S. and Moustakas, J. and Mueller, E. and Muñoz-Gutiérrez, A. and Myers, A. D. and Nadathur, S. and Napolitano, L. and Neveux, R. and Newman, J. A. and Nguyen, N. M. and Nie, J. and Niz, G. and Noriega, H. E. and Padmanabhan, N. and Paillas, E. and Palanque-Delabrouille, N. and Pan, J. and Penmetsa, S. and Percival, W. J. and Pieri, M. M. and Pinon, M. and Poppett, C. and Porredon, A. and Prada, F. and Pérez-Fernández, A. and Pérez-Ràfols, I. and Rabinowitz, D. and Raichoor, A. and Ramírez-Pérez, C. and Ramirez-Solano, S. and Rashkovetskyi, M. and Ravoux, C. and Rezaie, M. and Rich, J. and Rocher, A. and Rockosi, C. and Rodríguez-Martínez, F. and Roe, N. A. and Rosado-Marin, A. and Ross, A. J. and Rossi, G. and Ruggeri, R. and Ruhlmann-Kleider, V. and Samushia, L. and Sanchez, E. and Saulder, C. and Schlafly, E. F. and Schlegel, D. and Schubnell, M. and Seo, H. and Sharples, R. and Silber, J. and Slosar, A. and Smith, A. and Sprayberry, D. and Tan, T. and Tarlé, G. and Trusov, S. and Vaisakh, R. and Valcin, D. and Valdes, F. and Vargas-Magaña, M. and Verde, L. and Walther, M. and Wang, B. and Wang, M. S. and Weaver, B. A. and Weaverdyck, N. and Wechsler, R. H. and Weinberg, D. H. and White, M. and Wilson, M. J. and Yu, J. and Yu, Y. and Yuan, S. and Yèche, C. and Zaborowski, E. A. and Zarrouk, P. and Zhang, H. and Zhao, C. and Zhao, R. and Zhou, R. and Zou, H.},
	month = sep,
	year = {2025},
	note = {arXiv:2411.12021 [astro-ph]},
	pages = {008},
}

@article{garcia-quintero_hod-dependent_2025,
	title = {{HOD}-{Dependent} {Systematics} in {Emission} {Line} {Galaxies} for the {DESI} 2024 {BAO} analysis},
	volume = {2025},
	issn = {1475-7516},
	url = {http://arxiv.org/abs/2404.03009},
	doi = {10.1088/1475-7516/2025/01/132},
	language = {en},
	number = {01},
	urldate = {2025-11-03},
	journal = {Journal of Cosmology and Astroparticle Physics},
	author = {Garcia-Quintero, C. and Mena-Fernández, J. and Rocher, A. and Yuan, S. and Hadzhiyska, B. and Alves, O. and Rashkovetskyi, M. and Seo, H. and Padmanabhan, N. and Nadathur, S. and Howlett, C. and Ishak, M. and Medina-Varela, L. and McDonald, P. and Ross, A. J. and Xie, Y. and Chen, X. and Bera, A. and Aguilar, J. and Ahlen, S. and Andrade, U. and BenZvi, S. and Brooks, D. and Burtin, E. and Chen, S. and Claybaugh, T. and Cole, S. and Macorra, A. de la and Mattia, A. de and Dey, A. and Dey, B. and Ding, Z. and Doel, P. and Fanning, K. and Forero-Romero, J. E. and Gaztañaga, E. and Gil-Marín, H. and Gontcho, S. Gontcho A. and Gutierrez, G. and Guy, J. and Hahn, C. and Honscheid, K. and Kremin, A. and Landriau, M. and Guillou, L. Le and Levi, M. E. and Manera, M. and Martini, P. and Meisner, A. and Miquel, R. and Moustakas, J. and Mueller, E. and Muñoz-Gutiérrez, A. and Myers, A. D. and Newman, J. A. and Nie, J. and Niz, G. and Paillas, E. and Palanque-Delabrouille, N. and Percival, W. J. and Poppett, C. and Pérez-Fernández, A. and Rosado-Marin, A. and Rossi, G. and Ruggeri, R. and Sanchez, E. and Schlegel, D. and Schubnell, M. and Sprayberry, D. and Tarlé, G. and Vargas-Magaña, M. and Weaver, B. A. and Yu, J. and Zhang, H. and Zhou, R. and Zou, H.},
	month = jan,
	year = {2025},
	note = {arXiv:2404.03009 [astro-ph]},
	pages = {132},
}

@article{collaboration_desi_2025-2,
	title = {{DESI} 2024 {VI}: {Cosmological} {Constraints} from the {Measurements} of {Baryon} {Acoustic} {Oscillations}},
	volume = {2025},
	issn = {1475-7516},
	shorttitle = {{DESI} 2024 {VI}},
	url = {http://arxiv.org/abs/2404.03002},
	doi = {10.1088/1475-7516/2025/02/021},
	language = {en},
	number = {02},
	urldate = {2025-11-03},
	journal = {Journal of Cosmology and Astroparticle Physics},
	author = {Collaboration, DESI and Adame, A. G. and Aguilar, J. and Ahlen, S. and Alam, S. and Alexander, D. M. and Alvarez, M. and Alves, O. and Anand, A. and Andrade, U. and Armengaud, E. and Avila, S. and Aviles, A. and Awan, H. and Bahr-Kalus, B. and Bailey, S. and Baltay, C. and Bault, A. and Behera, J. and BenZvi, S. and Bera, A. and Beutler, F. and Bianchi, D. and Blake, C. and Blum, R. and Brieden, S. and Brodzeller, A. and Brooks, D. and Buckley-Geer, E. and Burtin, E. and Calderon, R. and Canning, R. and Rosell, A. Carnero and Cereskaite, R. and Cervantes-Cota, J. L. and Chabanier, S. and Chaussidon, E. and Chaves-Montero, J. and Chen, S. and Chen, X. and Claybaugh, T. and Cole, S. and Cuceu, A. and Davis, T. M. and Dawson, K. and Macorra, A. de la and Mattia, A. de and Deiosso, N. and Dey, A. and Dey, B. and Ding, Z. and Doel, P. and Edelstein, J. and Eftekharzadeh, S. and Eisenstein, D. J. and Elliott, A. and Fagrelius, P. and Fanning, K. and Ferraro, S. and Ereza, J. and Findlay, N. and Flaugher, B. and Font-Ribera, A. and Forero-Sánchez, D. and Forero-Romero, J. E. and Frenk, C. S. and Garcia-Quintero, C. and Gaztañaga, E. and Gil-Marín, H. and Gontcho, S. Gontcho A. and Gonzalez-Morales, A. X. and Gonzalez-Perez, V. and Gordon, C. and Green, D. and Gruen, D. and Gsponer, R. and Gutierrez, G. and Guy, J. and Hadzhiyska, B. and Hahn, C. and Hanif, M. M. S. and Herrera-Alcantar, H. K. and Honscheid, K. and Howlett, C. and Huterer, D. and Iršič, V. and Ishak, M. and Juneau, S. and Karaçaylı, N. G. and Kehoe, R. and Kent, S. and Kirkby, D. and Kremin, A. and Krolewski, A. and Lai, Y. and Lan, T.-W. and Landriau, M. and Lang, D. and Lasker, J. and Goff, J. M. Le and Guillou, L. Le and Leauthaud, A. and Levi, M. E. and Li, T. S. and Linder, E. and Lodha, K. and Magneville, C. and Manera, M. and Margala, D. and Martini, P. and Maus, M. and McDonald, P. and Medina-Varela, L. and Meisner, A. and Mena-Fernández, J. and Miquel, R. and Moon, J. and Moore, S. and Moustakas, J. and Mudur, N. and Mueller, E. and Muñoz-Gutiérrez, A. and Myers, A. D. and Nadathur, S. and Napolitano, L. and Neveux, R. and Newman, J. A. and Nguyen, N. M. and Nie, J. and Niz, G. and Noriega, H. E. and Padmanabhan, N. and Paillas, E. and Palanque-Delabrouille, N. and Pan, J. and Penmetsa, S. and Percival, W. J. and Pieri, M. M. and Pinon, M. and Poppett, C. and Porredon, A. and Prada, F. and Pérez-Fernández, A. and Pérez-Ràfols, I. and Rabinowitz, D. and Raichoor, A. and Ramírez-Pérez, C. and Ramirez-Solano, S. and Ravoux, C. and Rashkovetskyi, M. and Rezaie, M. and Rich, J. and Rocher, A. and Rockosi, C. and Roe, N. A. and Rosado-Marin, A. and Ross, A. J. and Rossi, G. and Ruggeri, R. and Ruhlmann-Kleider, V. and Samushia, L. and Sanchez, E. and Saulder, C. and Schlafly, E. F. and Schlegel, D. and Schubnell, M. and Seo, H. and Shafieloo, A. and Sharples, R. and Silber, J. and Slosar, A. and Smith, A. and Sprayberry, D. and Tan, T. and Tarlé, G. and Taylor, P. and Trusov, S. and Ureña-López, L. A. and Vaisakh, R. and Valcin, D. and Valdes, F. and Vargas-Magaña, M. and Verde, L. and Walther, M. and Wang, B. and Wang, M. S. and Weaver, B. A. and Weaverdyck, N. and Wechsler, R. H. and Weinberg, D. H. and White, M. and Yu, J. and Yu, Y. and Yuan, S. and Yèche, C. and Zaborowski, E. A. and Zarrouk, P. and Zhang, H. and Zhao, C. and Zhao, R. and Zhou, R. and Zhuang, T. and Zou, H.},
	month = feb,
	year = {2025},
	note = {arXiv:2404.03002 [astro-ph]},
	pages = {021},
}

@article{rocher_desi_2023,
	title = {The {DESI} {One}-{Percent} survey: exploring the {Halo} {Occupation} {Distribution} of {Emission} {Line} {Galaxies} with {AbacusSummit} simulations},
	volume = {2023},
	issn = {1475-7516},
	shorttitle = {The {DESI} {One}-{Percent} survey},
	url = {https://doi.org/10.1088/1475-7516/2023/10/016},
	doi = {10.1088/1475-7516/2023/10/016},
	language = {en},
	number = {10},
	urldate = {2026-09-16},
	journal = {Journal of Cosmology and Astroparticle Physics},
	publisher = {IOP Publishing},
	author = {Rocher, Antoine and Ruhlmann-Kleider, Vanina and Burtin, Etienne and Yuan, Sihan and de Mattia, Arnaud and Ross, Ashley J. and Aguilar, Jessica and Ahlen, Steven and Alam, Shadab and Bianchi, Davide and Brooks, David and Cole, Shaun and Dawson, Kyle and de la Macorra, Axel and Doel, Peter and Eisenstein, Daniel J. and Fanning, Kevin and Forero-Romero, Jaime E. and Garrison, Lehman H. and Gontcho A Gontcho, Satya and Gonzalez-Perez, Violeta and Guy, Julien and Hadzhiyska, Boryana and Hahn, ChangHoon and Honscheid, Klaus and Kisner, Theodore and Landriau, Martin and Lasker, James and E. Levi, Michael and Manera, Marc and Meisner, Aaron and Miquel, Ramon and Moustakas, John and Mueller, Eva-Maria and Newman, Jeffrey A. and Nie, Jundan and Percival, Will J. and Poppett, Claire and Qin, Fei and Rossi, Graziano and Samushia, Lado and Sanchez, Eusebio and Schlegel, David and Schubnell, Michael and Seo, Hee-Jong and Tarlé, Gregory and Vargas-Magaña, Mariana and Weaver, Benjamin A. and Yu, Jiaxi and Zhang, Hanyu and Zheng, Zheng and Zhou, Zhimin and Zou, Hu},
	month = oct,
	year = {2023},
	pages = {016},
}

@article{wechsler_connection_2018,
	title = {The {Connection} between {Galaxies} and their {Dark} {Matter} {Halos}},
	volume = {56},
	issn = {0066-4146, 1545-4282},
	url = {http://arxiv.org/abs/1804.03097},
	doi = {10.1146/annurev-astro-081817-051756},
	number = {1},
	urldate = {2025-11-03},
	journal = {Annual Review of Astronomy and Astrophysics},
	author = {Wechsler, Risa H. and Tinker, Jeremy L.},
	month = sep,
	year = {2018},
	note = {arXiv:1804.03097 [astro-ph]},
	pages = {435--487},
}

@article{ivezic_lsst_2019,
	title = {{LSST}: from {Science} {Drivers} to {Reference} {Design} and {Anticipated} {Data} {Products}},
	volume = {873},
	issn = {0004-637X, 1538-4357},
	shorttitle = {{LSST}},
	url = {http://arxiv.org/abs/0805.2366},
	doi = {10.3847/1538-4357/ab042c},
	language = {en},
	number = {2},
	urldate = {2025-11-04},
	journal = {The Astrophysical Journal},
	author = {Ivezić, {\v{Z}}eljko and Kahn, Steven M. and Tyson, J. Anthony and Abel, Bob and Acosta, Emily and Allsman, Robyn and Alonso, David and AlSayyad, Yusra and Anderson, Scott F. and Andrew, John and Angel, James Roger P. and Angeli, George Z. and Ansari, Reza and Antilogus, Pierre and Araujo, Constanza and Armstrong, Robert and Arndt, Kirk T. and Astier, Pierre and Aubourg, Éric and Auza, Nicole and Axelrod, Tim S. and Bard, Deborah J. and Barr, Jeff D. and Barrau, Aurelian and Bartlett, James G. and Bauer, Amanda E. and Bauman, Brian J. and Baumont, Sylvain and Becker, Andrew C. and Becla, Jacek and Beldica, Cristina and Bellavia, Steve and Bianco, Federica B. and Biswas, Rahul and Blanc, Guillaume and Blazek, Jonathan and Blandford, Roger D. and Bloom, Josh S. and Bogart, Joanne and Bond, Tim W. and Borgland, Anders W. and Borne, Kirk and Bosch, James F. and Boutigny, Dominique and Brackett, Craig A. and Bradshaw, Andrew and Brandt, William Nielsen and Brown, Michael E. and Bullock, James S. and Burchat, Patricia and Burke, David L. and Cagnoli, Gianpietro and Calabrese, Daniel and Callahan, Shawn and Callen, Alice L. and Chandrasekharan, Srinivasan and Charles-Emerson, Glenaver and Chesley, Steve and Cheu, Elliott C. and Chiang, Hsin-Fang and Chiang, James and Chirino, Carol and Chow, Derek and Ciardi, David R. and Claver, Charles F. and Cohen-Tanugi, Johann and Cockrum, Joseph J. and Coles, Rebecca and Connolly, Andrew J. and Cook, Kem H. and Cooray, Asantha and Covey, Kevin R. and Cribbs, Chris and Cui, Wei and Cutri, Roc and Daly, Philip N. and Daniel, Scott F. and Daruich, Felipe and Daubard, Guillaume and Daues, Greg and Dawson, William and Delgado, Francisco and Dellapenna, Alfred and Peyster, Robert de and Val-Borro, Miguel de and Digel, Seth W. and Doherty, Peter and Dubois, Richard and Dubois-Felsmann, Gregory P. and Durech, Josef and Economou, Frossie and Eracleous, Michael and Ferguson, Henry and Figueroa, Enrique and Fisher-Levine, Merlin and Focke, Warren and Foss, Michael D. and Frank, James and Freemon, Michael D. and Gangler, Emmanuel and Gawiser, Eric and Geary, John C. and Gee, Perry and Geha, Marla and Gessner, Charles J. B. and Gibson, Robert R. and Gilmore, D. Kirk and Glanzman, Thomas and Glick, William and Goldina, Tatiana and Goldstein, Daniel A. and Goodenow, Iain and Graham, Melissa L. and Gressler, William J. and Gris, Philippe and Guy, Leanne P. and Guyonnet, Augustin and Haller, Gunther and Harris, Ron and Hascall, Patrick A. and Haupt, Justine and Hernandez, Fabio and Herrmann, Sven and Hileman, Edward and Hoblitt, Joshua and Hodgson, John A. and Hogan, Craig and Huang, Dajun and Huffer, Michael E. and Ingraham, Patrick and Innes, Walter R. and Jacoby, Suzanne H. and Jain, Bhuvnesh and Jammes, Fabrice and Jee, James and Jenness, Tim and Jernigan, Garrett and Jevremović, Darko and Johns, Kenneth and Johnson, Anthony S. and Johnson, Margaret W. G. and Jones, R. Lynne and Juramy-Gilles, Claire and Jurić, Mario and Kalirai, Jason S. and Kallivayalil, Nitya J. and Kalmbach, Bryce and Kantor, Jeffrey P. and Karst, Pierre and Kasliwal, Mansi M. and Kelly, Heather and Kessler, Richard and Kinnison, Veronica and Kirkby, David and Knox, Lloyd and Kotov, Ivan V. and Krabbendam, Victor L. and Krughoff, K. Simon and Kubánek, Petr and Kuczewski, John and Kulkarni, Shri and Ku, John and Kurita, Nadine R. and Lage, Craig S. and Lambert, Ron and Lange, Travis and Langton, J. Brian and Guillou, Laurent Le and Levine, Deborah and Liang, Ming and Lim, Kian-Tat and Lintott, Chris J. and Long, Kevin E. and Lopez, Margaux and Lotz, Paul J. and Lupton, Robert H. and Lust, Nate B. and MacArthur, Lauren A. and Mahabal, Ashish and Mandelbaum, Rachel and Marsh, Darren S. and Marshall, Philip J. and Marshall, Stuart and May, Morgan and McKercher, Robert and McQueen, Michelle and Meyers, Joshua and Migliore, Myriam and Miller, Michelle and Mills, David J. and Miraval, Connor and Moeyens, Joachim and Monet, David G. and Moniez, Marc and Monkewitz, Serge and Montgomery, Christopher and Mueller, Fritz and Muller, Gary P. and Arancibia, Freddy Muñoz and Neill, Douglas R. and Newbry, Scott P. and Nief, Jean-Yves and Nomerotski, Andrei and Nordby, Martin and O'Connor, Paul and Oliver, John and Olivier, Scot S. and Olsen, Knut and O'Mullane, William and Ortiz, Sandra and Osier, Shawn and Owen, Russell E. and Pain, Reynald and Palecek, Paul E. and Parejko, John K. and Parsons, James B. and Pease, Nathan M. and Peterson, J. Matt and Peterson, John R. and Petravick, Donald L. and Petrick, M. E. Libby and Petry, Cathy E. and Pierfederici, Francesco and Pietrowicz, Stephen and Pike, Rob and Pinto, Philip A. and Plante, Raymond and Plate, Stephen and Price, Paul A. and Prouza, Michael and Radeka, Veljko and Rajagopal, Jayadev and Rasmussen, Andrew P. and Regnault, Nicolas and Reil, Kevin A. and Reiss, David J. and Reuter, Michael A. and Ridgway, Stephen T. and Riot, Vincent J. and Ritz, Steve and Robinson, Sean and Roby, William and Roodman, Aaron and Rosing, Wayne and Roucelle, Cecille and Rumore, Matthew R. and Russo, Stefano and Saha, Abhijit and Sassolas, Benoit and Schalk, Terry L. and Schellart, Pim and Schindler, Rafe H. and Schmidt, Samuel and Schneider, Donald P. and Schneider, Michael D. and Schoening, William and Schumacher, German and Schwamb, Megan E. and Sebag, Jacques and Selvy, Brian and Sembroski, Glenn H. and Seppala, Lynn G. and Serio, Andrew and Serrano, Eduardo and Shaw, Richard A. and Shipsey, Ian and Sick, Jonathan and Silvestri, Nicole and Slater, Colin T. and Smith, J. Allyn and Smith, R. Chris and Sobhani, Shahram and Soldahl, Christine and Storrie-Lombardi, Lisa and Stover, Edward and Strauss, Michael A. and Street, Rachel A. and Stubbs, Christopher W. and Sullivan, Ian S. and Sweeney, Donald and Swinbank, John D. and Szalay, Alexander and Takacs, Peter and Tether, Stephen A. and Thaler, Jon J. and Thayer, John Gregg and Thomas, Sandrine and Thukral, Vaikunth and Tice, Jeffrey and Trilling, David E. and Turri, Max and Berg, Richard Van and Berk, Daniel Vanden and Vetter, Kurt and Virieux, Francoise and Vucina, Tomislav and Wahl, William and Walkowicz, Lucianne and Walsh, Brian and Walter, Christopher W. and Wang, Daniel L. and Wang, Shin-Yawn and Warner, Michael and Wiecha, Oliver and Willman, Beth and Winters, Scott E. and Wittman, David and Wolff, Sidney C. and Wood-Vasey, W. Michael and Wu, Xiuqin and Xin, Bo and Yoachim, Peter and Zhan, Hu},
	month = mar,
	year = {2019},
	note = {arXiv:0805.2366 [astro-ph]},
	pages = {111},
}

@misc{yuan_desi_2023,
	title = {The {DESI} {One}-{Percent} {Survey}: {Exploring} the {Halo} {Occupation} {Distribution} of {Luminous} {Red} {Galaxies} and {Quasi}-{Stellar} {Objects} with {AbacusSummit}},
	shorttitle = {The {DESI} {One}-{Percent} {Survey}},
	url = {http://arxiv.org/abs/2306.06314},
	doi = {10.48550/arXiv.2306.06314},
	urldate = {2025-11-04},
	publisher = {arXiv},
	author = {Yuan, Sihan and Zhang, Hanyu and Ross, Ashley J. and Donald-McCann, Jamie and Hadzhiyska, Boryana and Wechsler, Risa H. and Zheng, Zheng and Alam, Shadab and Gonzalez-Perez, Violeta and Aguilar, Jessica Nicole and Ahlen, Steven and Bianchi, Davide and Brooks, David and Macorra, Axel de la and Fanning, Kevin and Forero-Romero, Jaime E. and Honscheid, Klaus and Ishak, Mustapha and Kehoe, Robert and Lasker, James and Landriau, Martin and Manera, Marc and Martini, Paul and Meisner, Aaron and Miquel, Ramon and Moustakas, John and Nadathur, Seshadri and Newman, Jeffrey A. and Nie, Jundan and Percival, Will and Poppett, Claire and Rocher, Antoine and Rossi, Graziano and Sanchez, Eusebio and Samushia, Lado and Schubnell, Michael and Seo, Hee-Jong and Tarle, Gregory and Weaver, Benjamin Alan and Yu, Jiaxi and Zhou, Zhimin and Zou, Hu},
	month = jun,
	year = {2023},
	note = {arXiv:2306.06314 [astro-ph]},
}

@misc{wang_galaxy-multiplet_2025,
	title = {Galaxy-{Multiplet} {Clustering} from {DESI} {DR2}},
	url = {http://arxiv.org/abs/2511.15354},
	doi = {10.48550/arXiv.2511.15354},
	language = {en},
	urldate = {2025-11-26},
	publisher = {arXiv},
	author = {Wang, Hanyue and Eisenstein, Daniel J. and Aguilar, Jessica Nicole and Ahlen, Steven and Bianchi, Davide and Brooks, David and Claybaugh, Todd and Macorra, Axel de la and Dey, Arjun and Dey, Biprateep and Doel, Peter and Ferraro, Simone and Font-Ribera, Andreu and Forero-Romero, Jaime E. and Gaztañaga, Enrique and Gutierrez, Gaston and Honscheid, Klaus and Ishak, Mustapha and Joyce, Richard and Juneau, Stephanie and Kirkby, David and Kisner, Theodore and Kremin, Anthony and Lahav, Ofer and Lamman, Claire and Landriau, Martin and Manera, Marc and Meisner, Aaron and Miquel, Ramon and Mueller, Eva-Maria and Nadathur, Seshadri and Niz, Gustavo and Palanque-Delabrouille, Nathalie and Percival, Will J. and Prada, Francisco and Pérez-Ràfols, Ignasi and Ross, Ashley J. and Rossi, Graziano and Sanchez, Eusebio and Schlegel, David and Schubnell, Michael and Silber, Joseph Harry and Sprayberry, David and Tarlé, Gregory and Weaver, Benjamin Alan and Zhou, Rongpu and Zou, Hu},
	month = nov,
	year = {2025},
	note = {arXiv:2511.15354 [astro-ph]},
}

@article{gaunt_asymptotic_2019,
	title = {An asymptotic expansion for the normalizing constant of the {Conway}–{Maxwell}–{Poisson} distribution},
	volume = {71},
	issn = {1572-9052},
	url = {https://doi.org/10.1007/s10463-017-0629-6},
	doi = {10.1007/s10463-017-0629-6},
	language = {en},
	number = {1},
	urldate = {2026-02-18},
	journal = {Annals of the Institute of Statistical Mathematics},
	author = {Gaunt, Robert E. and Iyengar, Satish and Olde Daalhuis, Adri B. and Simsek, Burcin},
	month = feb,
	year = {2019},
	pages = {163--180},
}

@article{chaves-montero_galaxy_2023,
	title = {The galaxy formation origin of the lensing is low problem},
	volume = {521},
	issn = {0035-8711, 1365-2966},
	url = {http://arxiv.org/abs/2211.01744},
	doi = {10.1093/mnras/stad243},
	language = {en},
	number = {1},
	urldate = {2026-03-05},
	journal = {Monthly Notices of the Royal Astronomical Society},
	author = {Chaves-Montero, Jonas and Angulo, Raul E. and Contreras, Sergio},
	month = mar,
	year = {2023},
	note = {arXiv:2211.01744 [astro-ph]},
	pages = {937--951},
}

@article{vos-gines_improving_2024,
	title = {Improving and extending non-{Poissonian} distributions for satellite galaxies sampling in {HOD}: applications to {eBOSS} {ELGs}},
	volume = {530},
	copyright = {https://creativecommons.org/licenses/by/4.0/},
	issn = {0035-8711, 1365-2966},
	shorttitle = {Improving and extending non-{Poissonian} distributions for satellite galaxies sampling in {HOD}},
	url = {https://academic.oup.com/mnras/article/530/3/3458/7656975},
	doi = {10.1093/mnras/stae1096},
	language = {en},
	number = {3},
	urldate = {2026-09-16},
	journal = {Monthly Notices of the Royal Astronomical Society},
	author = {Vos-Ginés, Bernhard and Avila, Santiago and Gonzalez-Perez, Violeta and Yepes, Gustavo},
	month = apr,
	year = {2024},
	pages = {3458--3476},
}

@article{avila_completed_2020,
	title = {The {Completed} {SDSS}-{IV} extended {Baryon} {Oscillation} {Spectroscopic} {Survey}: exploring the {Halo} {Occupation} {Distribution} model for {Emission} {Line} {Galaxies}},
	volume = {499},
	issn = {0035-8711, 1365-2966},
	shorttitle = {The {Completed} {SDSS}-{IV} extended {Baryon} {Oscillation} {Spectroscopic} {Survey}},
	url = {http://arxiv.org/abs/2007.09012},
	doi = {10.1093/mnras/staa2951},
	language = {en},
	number = {4},
	urldate = {2026-03-05},
	journal = {Monthly Notices of the Royal Astronomical Society},
	author = {Avila, Santiago and Gonzalez-Perez, Violeta and Mohammad, Faizan G. and Mattia, Arnaud de and Zhao, Cheng and Raichoor, Anand and Tamone, Amelie and Alam, Shadab and Bautista, Julian and Bianchi, Davide and Burtin, Etienne and Chapman, Michael J. and Chuang, Chia-Hsun and Comparat, Johan and Dawson, Kyle and Divers, Thomas and Bourboux, Helion du Mas des and Gil-Marin, Hector and Mueller, Eva-Maria and Habib, Salman and Heitmann, Katrin and Ruhlmann-Kleider, Vanina and Padilla, Nelson and Percival, Will J. and Ross, Ashley J. and Seo, Hee-Jong and Schneider, Donald P. and Zhao, Gong-Bo},
	month = nov,
	year = {2020},
	note = {arXiv:2007.09012 [astro-ph]},
	pages = {5486--5507},
}

@article{jimenez_extensions_2019,
	title = {Extensions to the halo occupation distribution model for more accurate clustering predictions},
	volume = {490},
	copyright = {https://academic.oup.com/journals/pages/open\_access/funder\_policies/chorus/standard\_publication\_model},
	issn = {0035-8711, 1365-2966},
	url = {https://academic.oup.com/mnras/article/490/3/3532/5581510},
	doi = {10.1093/mnras/stz2790},
	language = {en},
	number = {3},
	urldate = {2026-03-05},
	journal = {Monthly Notices of the Royal Astronomical Society},
	author = {Jiménez, Esteban and Contreras, Sergio and Padilla, Nelson and Zehavi, Idit and Baugh, Carlton M and Gonzalez-Perez, Violeta},
	month = dec,
	year = {2019},
	pages = {3532--3544},
}

@article{boylan-kolchin_theres_2010,
	title = {There's no place like home? {Statistics} of {Milky} {Way}-mass dark matter haloes: {Statistics} of {Milky} {Way}-mass haloes},
	issn = {00358711, 13652966},
	shorttitle = {There's no place like home?},
	url = {https://academic.oup.com/mnras/article-lookup/doi/10.1111/j.1365-2966.2010.16774.x},
	doi = {10.1111/j.1365-2966.2010.16774.x},
	language = {en},
	urldate = {2026-03-05},
	journal = {Monthly Notices of the Royal Astronomical Society},
	author = {Boylan-Kolchin, Michael and Springel, Volker and White, Simon D. M. and Jenkins, Adrian},
	month = jun,
	year = {2010},
	pages = {no--no},
}

@article{jiang_statistics_2017,
	title = {Statistics of dark matter substructure – {III}. {Halo}-to-halo variance},
	volume = {472},
	issn = {0035-8711, 1365-2966},
	url = {http://academic.oup.com/mnras/article/472/1/657/4064375/Statistics-of-dark-matter-substructure-III},
	doi = {10.1093/mnras/stx1979},
	language = {en},
	number = {1},
	urldate = {2026-03-05},
	journal = {Monthly Notices of the Royal Astronomical Society},
	author = {Jiang, Fangzhou and Van Den Bosch, Frank C.},
	month = nov,
	year = {2017},
	pages = {657--674},
}

@article{hadzhiyska_millenniumtng_2023,
	title = {The {MillenniumTNG} {Project}: {Refining} the one-halo model of red and blue galaxies at different redshifts},
	volume = {524},
	issn = {0035-8711, 1365-2966},
	shorttitle = {The {MillenniumTNG} {Project}},
	url = {http://arxiv.org/abs/2210.10068},
	doi = {10.1093/mnras/stad279},
	language = {en},
	number = {2},
	urldate = {2026-03-05},
	journal = {Monthly Notices of the Royal Astronomical Society},
	author = {Hadzhiyska, Boryana and Hernquist, Lars and Eisenstein, Daniel and Delgado, Ana Maria and Bose, Sownak and Kannan, Rahul and Pakmor, Rüdiger and Springel, Volker and Contreras, Sergio and Barrera, Monica and Ferlito, Fulvio and Hernández-Aguayo, César and White, Simon D. M. and Frenk, Carlos},
	month = jul,
	year = {2023},
	note = {arXiv:2210.10068 [astro-ph]},
	pages = {2524--2538},
}

@article{reid_constraining_2009,
	title = {Constraining the {LRG} {Halo} {Occupation} {Distribution} using {Counts}-in-{Cylinders}},
	volume = {698},
	issn = {0004-637X, 1538-4357},
	url = {http://arxiv.org/abs/0809.4505},
	doi = {10.1088/0004-637X/698/1/143},
	number = {1},
	urldate = {2026-03-20},
	journal = {The Astrophysical Journal},
	author = {Reid, Beth A. and Spergel, David N.},
	month = jun,
	year = {2009},
	note = {arXiv:0809.4505 [astro-ph]},
	pages = {143--154},
}

@article{zheng_galaxy_2007,
	title = {Galaxy {Evolution} from {Halo} {Occupation} {Distribution} {Modeling} of {DEEP2} and {SDSS} {Galaxy} {Clustering}},
	volume = {667},
	issn = {0004-637X},
	url = {https://iopscience.iop.org/article/10.1086/521074},
	doi = {10.1086/521074},
	language = {en},
	number = {2},
	urldate = {2026-03-23},
	journal = {The Astrophysical Journal},
	publisher = {IOP Publishing},
	author = {Zheng, Zheng and Coil, Alison L. and Zehavi, Idit},
	month = oct,
	year = {2007},
	pages = {760},
}

@article{zheng_theoretical_2005,
	title = {Theoretical {Models} of the {Halo} {Occupation} {Distribution}: {Separating} {Central} and {Satellite} {Galaxies}},
	volume = {633},
	issn = {0004-637X, 1538-4357},
	shorttitle = {Theoretical {Models} of the {Halo} {Occupation} {Distribution}},
	url = {http://arxiv.org/abs/astro-ph/0408564},
	doi = {10.1086/466510},
	language = {en},
	number = {2},
	urldate = {2026-03-23},
	journal = {The Astrophysical Journal},
	author = {Zheng, Zheng and Berlind, Andreas A. and Weinberg, David H. and Benson, Andrew J. and Baugh, Carlton M. and Cole, Shaun and Dave, Romeel and Frenk, Carlos S. and Katz, Neal and Lacey, Cedric G.},
	month = nov,
	year = {2005},
	note = {arXiv:astro-ph/0408564},
	pages = {791--809},
}

@misc{cabass_snowmass_2022,
	title = {Snowmass {White} {Paper}: {Effective} {Field} {Theories} in {Cosmology}},
	shorttitle = {Snowmass {White} {Paper}},
	url = {http://arxiv.org/abs/2203.08232},
	doi = {10.48550/arXiv.2203.08232},
	urldate = {2026-03-23},
	publisher = {arXiv},
	author = {Cabass, Giovanni and Ivanov, Mikhail M. and Lewandowski, Matthew and Mirbabayi, Mehrdad and Simonović, Marko},
	month = mar,
	year = {2022},
	note = {arXiv:2203.08232 [astro-ph]},
}

@article{senatore_bias_2015,
	title = {Bias in the {Effective} {Field} {Theory} of {Large} {Scale} {Structures}},
	volume = {2015},
	issn = {1475-7516},
	url = {http://arxiv.org/abs/1406.7843},
	doi = {10.1088/1475-7516/2015/11/007},
	number = {11},
	urldate = {2026-03-23},
	journal = {Journal of Cosmology and Astroparticle Physics},
	author = {Senatore, Leonardo},
	month = nov,
	year = {2015},
	note = {arXiv:1406.7843 [astro-ph]},
	pages = {007--007},
}

@article{white_core_1978,
	title = {Core condensation in heavy halos: a two-stage theory for galaxy formation and clustering.},
	volume = {183},
	issn = {0035-8711},
	shorttitle = {Core condensation in heavy halos},
	url = {https://ui.adsabs.harvard.edu/abs/1978MNRAS.183..341W},
	doi = {10.1093/mnras/183.3.341},
	urldate = {2026-03-23},
	journal = {Monthly Notices of the Royal Astronomical Society},
	publisher = {OUP},
	author = {White, S. D. M. and Rees, M. J.},
	month = may,
	year = {1978},
	note = {ADS Bibcode: 1978MNRAS.183..341W},
	pages = {341--358},
}

@article{cole_hierarchical_2000,
	title = {Hierarchical galaxy formation},
	volume = {319},
	issn = {0035-8711},
	url = {https://ui.adsabs.harvard.edu/abs/2000MNRAS.319..168C},
	doi = {10.1046/j.1365-8711.2000.03879.x},
	urldate = {2026-03-23},
	journal = {Monthly Notices of the Royal Astronomical Society},
	publisher = {OUP},
	author = {Cole, Shaun and Lacey, Cedric G. and Baugh, Carlton M. and Frenk, Carlos S.},
	month = nov,
	year = {2000},
	note = {ADS Bibcode: 2000MNRAS.319..168C},
	pages = {168--204},
}

@article{baugh_primer_2006,
	title = {A primer on hierarchical galaxy formation: the semi-analytical approach},
	volume = {69},
	issn = {0034-4885, 1361-6633},
	shorttitle = {A primer on hierarchical galaxy formation},
	url = {http://arxiv.org/abs/astro-ph/0610031},
	doi = {10.1088/0034-4885/69/12/R02},
	number = {12},
	urldate = {2026-03-23},
	journal = {Reports on Progress in Physics},
	author = {Baugh, C. M.},
	month = dec,
	year = {2006},
	note = {arXiv:astro-ph/0610031},
	pages = {3101--3156},
}

@article{kravtsov_dark_2004,
	title = {The {Dark} {Side} of the {Halo} {Occupation} {Distribution}},
	volume = {609},
	issn = {0004-637X},
	url = {https://ui.adsabs.harvard.edu/abs/2004ApJ...609...35K},
	doi = {10.1086/420959},
	urldate = {2026-03-23},
	journal = {The Astrophysical Journal},
	publisher = {IOP},
	author = {Kravtsov, Andrey V. and Berlind, Andreas A. and Wechsler, Risa H. and Klypin, Anatoly A. and Gottlöber, Stefan and Allgood, Brandon and Primack, Joel R.},
	month = jul,
	year = {2004},
	note = {ADS Bibcode: 2004ApJ...609...35K},
	pages = {35--49},
}

@article{tasitsiomi_modeling_2004,
	title = {Modeling {Galaxy}-{Mass} {Correlations} in {Dissipationless} {Simulations}},
	volume = {614},
	issn = {0004-637X},
	url = {https://iopscience.iop.org/article/10.1086/423784},
	doi = {10.1086/423784},
	language = {en},
	number = {2},
	urldate = {2026-03-23},
	journal = {The Astrophysical Journal},
	publisher = {IOP Publishing},
	author = {Tasitsiomi, Argyro and Kravtsov, Andrey V. and Wechsler, Risa H. and Primack, Joel R.},
	month = oct,
	year = {2004},
	pages = {533},
}

@article{vale_non-parametric_2006,
	title = {The non-parametric model for linking galaxy luminosity with halo/subhalo mass},
	volume = {371},
	issn = {0035-8711},
	url = {https://ui.adsabs.harvard.edu/abs/2006MNRAS.371.1173V},
	doi = {10.1111/j.1365-2966.2006.10605.x},
	urldate = {2026-03-23},
	journal = {Monthly Notices of the Royal Astronomical Society},
	publisher = {OUP},
	author = {Vale, A. and Ostriker, J. P.},
	month = sep,
	year = {2006},
	note = {ADS Bibcode: 2006MNRAS.371.1173V},
	pages = {1173--1187},
}

@article{conroy_modeling_2006,
	title = {Modeling {Luminosity}-dependent {Galaxy} {Clustering} through {Cosmic} {Time}},
	volume = {647},
	issn = {0004-637X},
	url = {https://ui.adsabs.harvard.edu/abs/2006ApJ...647..201C},
	doi = {10.1086/503602},
	urldate = {2026-03-23},
	journal = {The Astrophysical Journal},
	publisher = {IOP},
	author = {Conroy, Charlie and Wechsler, Risa H. and Kravtsov, Andrey V.},
	month = aug,
	year = {2006},
	note = {ADS Bibcode: 2006ApJ...647..201C},
	pages = {201--214},
}

@article{favole_clustering_2016,
	title = {Clustering properties of g-selected galaxies at z \ensuremath{\sim} 0.8},
	volume = {461},
	issn = {0035-8711},
	url = {https://ui.adsabs.harvard.edu/abs/2016MNRAS.461.3421F},
	doi = {10.1093/mnras/stw1483},
	urldate = {2026-03-23},
	journal = {Monthly Notices of the Royal Astronomical Society},
	publisher = {OUP},
	author = {Favole, Ginevra and Comparat, Johan and Prada, Francisco and Yepes, Gustavo and Jullo, Eric and Niemiec, Anna and Kneib, Jean-Paul and Rodríguez-Torres, Sergio A. and Klypin, Anatoly and Skibba, Ramin A. and McBride, Cameron K. and Eisenstein, Daniel J. and Schlegel, David J. and Nuza, Sebastián E. and Chuang, Chia-Hsun and Delubac, Timothée and Yèche, Christophe and Schneider, Donald P.},
	month = oct,
	year = {2016},
	note = {ADS Bibcode: 2016MNRAS.461.3421F},
	pages = {3421--3431},
}

@article{navarro_universal_1997,
	title = {A {Universal} {Density} {Profile} from {Hierarchical} {Clustering}},
	volume = {490},
	issn = {0004-637X},
	url = {https://ui.adsabs.harvard.edu/abs/1997ApJ...490..493N},
	doi = {10.1086/304888},
	urldate = {2026-03-23},
	journal = {The Astrophysical Journal},
	publisher = {IOP},
	author = {Navarro, Julio F. and Frenk, Carlos S. and White, Simon D. M.},
	month = dec,
	year = {1997},
	note = {ADS Bibcode: 1997ApJ...490..493N},
	pages = {493--508},
}

@misc{nelson_illustristng_2021,
	title = {The {IllustrisTNG} {Simulations}: {Public} {Data} {Release}},
	shorttitle = {The {IllustrisTNG} {Simulations}},
	url = {http://arxiv.org/abs/1812.05609},
	doi = {10.48550/arXiv.1812.05609},
	urldate = {2026-03-23},
	publisher = {arXiv},
	author = {Nelson, Dylan and Springel, Volker and Pillepich, Annalisa and Rodriguez-Gomez, Vicente and Torrey, Paul and Genel, Shy and Vogelsberger, Mark and Pakmor, Ruediger and Marinacci, Federico and Weinberger, Rainer and Kelley, Luke and Lovell, Mark and Diemer, Benedikt and Hernquist, Lars},
	month = jan,
	year = {2021},
	note = {arXiv:1812.05609 [astro-ph]},
}

@misc{lu_preference_2025,
	title = {Preference for evolving dark energy in light of the galaxy bispectrum},
	url = {http://arxiv.org/abs/2503.04602},
	doi = {10.48550/arXiv.2503.04602},
	urldate = {2026-03-23},
	publisher = {arXiv},
	author = {Lu, Zhiyu and Simon, Théo and Zhang, Pierre},
	month = apr,
	year = {2025},
	note = {arXiv:2503.04602 [astro-ph]},
}

@article{damico_boss_2024,
	title = {The {BOSS} bispectrum analysis at one loop from the {Effective} {Field} {Theory} of {Large}-{Scale} {Structure}},
	volume = {2024},
	issn = {1475-7516},
	url = {http://arxiv.org/abs/2206.08327},
	doi = {10.1088/1475-7516/2024/05/059},
	number = {05},
	urldate = {2026-03-23},
	journal = {Journal of Cosmology and Astroparticle Physics},
	author = {D'Amico, Guido and Donath, Yaniv and Lewandowski, Matthew and Senatore, Leonardo and Zhang, Pierre},
	month = may,
	year = {2024},
	note = {arXiv:2206.08327 [astro-ph]},
	pages = {059},
}

@misc{noauthor_r_nodate,
	title = {R. {W}. {Conway} and {W}. {L}. {Maxwell}, “{A} {Queuing} {Model} with {State} {Dependent} {Service} {Rates},” {Journal} of {Industrial} {Engineering}, {Vol}. 12, 1962, pp. 132-136. - {References} - {Scientific} {Research} {Publishing}},
	url = {https://www.scirp.org/reference/referencespapers?referenceid=502472},
	urldate = {2026-03-23},
}

@article{zehavi_galaxy_2011,
	title = {Galaxy {Clustering} in the {Completed} {SDSS} {Redshift} {Survey}: {The} {Dependence} on {Color} and {Luminosity}},
	volume = {736},
	issn = {0004-637X},
	shorttitle = {Galaxy {Clustering} in the {Completed} {SDSS} {Redshift} {Survey}},
	url = {https://ui.adsabs.harvard.edu/abs/2011ApJ...736...59Z},
	doi = {10.1088/0004-637X/736/1/59},
	urldate = {2026-03-23},
	journal = {The Astrophysical Journal},
	publisher = {IOP},
	author = {Zehavi, Idit and Zheng, Zheng and Weinberg, David H. and Blanton, Michael R. and Bahcall, Neta A. and Berlind, Andreas A. and Brinkmann, Jon and Frieman, Joshua A. and Gunn, James E. and Lupton, Robert H. and Nichol, Robert C. and Percival, Will J. and Schneider, Donald P. and Skibba, Ramin A. and Strauss, Michael A. and Tegmark, Max and York, Donald G.},
	month = jul,
	year = {2011},
	note = {ADS Bibcode: 2011ApJ...736...59Z},
	pages = {59},
}

@article{smith_completed_2020,
	title = {The completed {SDSS}-{IV} extended {Baryon} {Oscillation} {Spectroscopic} {Survey}: {N}-body mock challenge for the quasar sample},
	volume = {499},
	issn = {0035-8711},
	shorttitle = {The completed {SDSS}-{IV} extended {Baryon} {Oscillation} {Spectroscopic} {Survey}},
	url = {https://doi.org/10.1093/mnras/staa2825},
	doi = {10.1093/mnras/staa2825},
	number = {1},
	urldate = {2026-03-23},
	journal = {Monthly Notices of the Royal Astronomical Society},
	author = {Smith, Alex and Burtin, Etienne and Hou, Jiamin and Neveux, Richard and Ross, Ashley J and Alam, Shadab and Brinkmann, Jonathan and Dawson, Kyle S and Habib, Salman and Heitmann, Katrin and Kneib, Jean-Paul and Lyke, Brad W and du Mas des Bourboux, Hélion and Mueller, Eva-Maria and Myers, Adam D and Percival, Will J and Rossi, Graziano and Schneider, Donald P and Zarrouk, Pauline and Zhao, Gong-Bo},
	month = oct,
	year = {2020},
	pages = {269--291},
}

@article{sinha_corrfunc_2020,
	title = {Corrfunc --- {A} {Suite} of {Blazing} {Fast} {Correlation} {Functions} on the {CPU}},
	volume = {491},
	issn = {0035-8711, 1365-2966},
	url = {http://arxiv.org/abs/1911.03545},
	doi = {10.1093/mnras/stz3157},
	number = {2},
	urldate = {2026-03-23},
	journal = {Monthly Notices of the Royal Astronomical Society},
	author = {Sinha, Manodeep and Garrison, Lehman H.},
	month = jan,
	year = {2020},
	note = {arXiv:1911.03545 [astro-ph]},
	pages = {3022--3041},
}

@misc{noauthor_peebles_nodate,
	title = {Peebles {Hauser} 1974 - {Recherche} {Google}},
	url = {https://www.google.com/search?client=ubuntu-sn&channel=fs&q=Peebles+Hauser+1974},
	urldate = {2026-03-23},
}

@article{asgari_halo_2023,
	title = {The halo model for cosmology: a pedagogical review},
	volume = {6},
	issn = {2565-6120},
	shorttitle = {The halo model for cosmology},
	url = {http://arxiv.org/abs/2303.08752},
	doi = {10.21105/astro.2303.08752},
	language = {en},
	urldate = {2026-03-23},
	journal = {The Open Journal of Astrophysics},
	author = {Asgari, Marika and Mead, Alexander J. and Heymans, Catherine},
	month = nov,
	year = {2023},
	note = {arXiv:2303.08752 [astro-ph]},
	pages = {10.21105/astro.2303.08752},
}

@article{seljak_analytic_2000,
	title = {Analytic model for galaxy and dark matter clustering},
	volume = {318},
	issn = {0035-8711},
	url = {https://doi.org/10.1046/j.1365-8711.2000.03715.x},
	doi = {10.1046/j.1365-8711.2000.03715.x},
	number = {1},
	urldate = {2026-03-23},
	journal = {Monthly Notices of the Royal Astronomical Society},
	author = {Seljak, Uroš},
	month = oct,
	year = {2000},
	pages = {203--213},
}

@article{wang_evidence_2022,
	title = {Evidence of galaxy assembly bias in {SDSS} {DR7} galaxy samples from count statistics},
	volume = {516},
	issn = {0035-8711},
	url = {https://doi.org/10.1093/mnras/stac2465},
	doi = {10.1093/mnras/stac2465},
	number = {3},
	urldate = {2026-03-23},
	journal = {Monthly Notices of the Royal Astronomical Society},
	author = {Wang, Kuan and Mao, Yao-Yuan and Zentner, Andrew R and Guo, Hong and Lange, Johannes U and van den Bosch, Frank C and Mezini, Lorena},
	month = nov,
	year = {2022},
	pages = {4003--4024},
}

@misc{pearl_desi_2023,
	title = {The {DESI} {One}-{Percent} {Survey}: {Evidence} for {Assembly} {Bias} from {Low}-{Redshift} {Counts}-in-{Cylinders} {Measurements}},
	shorttitle = {The {DESI} {One}-{Percent} {Survey}},
	url = {http://arxiv.org/abs/2309.08675},
	doi = {10.48550/arXiv.2309.08675},
	language = {en},
	urldate = {2026-03-23},
	publisher = {arXiv},
	author = {Pearl, Alan N. and Zentner, Andrew R. and Newman, Jeffrey A. and Bezanson, Rachel and Wang, Kuan and Moustakas, John and Aguilar, Jessica N. and Ahlen, Steven and Brooks, David and Claybaugh, Todd and Cole, Shaun and Dawson, Kyle and Macorra, Axel de la and Doel, Peter and Forero-Romero, Jamie E. and Gontcho, Satya Gontcho A. and Honscheid, Klaus and Landriau, Martin and Manera, Marc and Meisner, Paul Martini Aaron and Miquel, Ramon and Nie, Jundan and Percival, Will and Prada, Francisco and Rezaie, Mehdi and Rossi, Graziano and Sanchez, Eusebio and Schubnell, Michael and Tarle, Gregory and Weaver, Benjamin A. and Zhou, Zhimin},
	month = sep,
	year = {2023},
	note = {arXiv:2309.08675 [astro-ph]},
}

@article{sugiyama_complete_2019,
	title = {A complete {FFT}-based decomposition formalism for the redshift-space bispectrum},
	volume = {484},
	issn = {0035-8711, 1365-2966},
	url = {http://arxiv.org/abs/1803.02132},
	doi = {10.1093/mnras/sty3249},
	language = {en},
	number = {1},
	urldate = {2026-04-15},
	journal = {Monthly Notices of the Royal Astronomical Society},
	author = {Sugiyama, Naonori S. and Saito, Shun and Beutler, Florian and Seo, Hee-Jong},
	month = mar,
	year = {2019},
	note = {arXiv:1803.02132 [astro-ph]},
	pages = {364--384},
}

@article{alcock_evolution_1979,
	title = {An evolution free test for non-zero cosmological constant},
	volume = {281},
	issn = {0028-0836},
	url = {https://ui.adsabs.harvard.edu/abs/1979Natur.281..358A},
	doi = {10.1038/281358a0},
	urldate = {2026-04-15},
	journal = {Nature},
	author = {Alcock, C. and Paczynski, B.},
	month = oct,
	year = {1979},
	note = {ADS Bibcode: 1979Natur.281..358A},
	pages = {358},
}

@book{varshalovich_quantum_1988,
	title = {Quantum {Theory} of {Angular} {Momentum}},
	url = {https://ui.adsabs.harvard.edu/abs/1988qtam.book.....V},
	doi = {10.1142/0270},
	urldate = {2026-04-16},
	author = {Varshalovich, D. A. and Moskalev, A. N. and Khersonskii, V. K.},
	month = jan,
	year = {1988},
	note = {Publication Title: Quantum Theory of Angular Momentum. Edited by VARSHALOVICH D A
ADS Bibcode: 1988qtam.book.....V},
}

@article{collaboration_euclid_2025,
	title = {Euclid. {I}. {Overview} of the {Euclid} mission},
	volume = {697},
	issn = {0004-6361, 1432-0746},
	url = {http://arxiv.org/abs/2405.13491},
	doi = {10.1051/0004-6361/202450810},
	urldate = {2026-09-16},
	journal = {Astronomy \& Astrophysics},
	author = {Collaboration, Euclid and Mellier, Y. and Abdurro'uf and Barroso, J. A. Acevedo and Achúcarro, A. and Adamek, J. and Adam, R. and Addison, G. E. and Aghanim, N. and Aguena, M. and Ajani, V. and Akrami, Y. and Al-Bahlawan, A. and Alavi, A. and Albuquerque, I. S. and Alestas, G. and Alguero, G. and Allaoui, A. and Allen, S. W. and Allevato, V. and Alonso-Tetilla, A. V. and Altieri, B. and Alvarez-Candal, A. and Alvi, S. and Amara, A. and Amendola, L. and Amiaux, J. and Andika, I. T. and Andreon, S. and Andrews, A. and Angora, G. and Angulo, R. E. and Annibali, F. and Anselmi, A. and Anselmi, S. and Arcari, S. and Archidiacono, M. and Aricò, G. and Arnaud, M. and Arnouts, S. and Asgari, M. and Asorey, J. and Atayde, L. and Atek, H. and Atrio-Barandela, F. and Aubert, M. and Aubourg, E. and Auphan, T. and Auricchio, N. and Aussel, B. and Aussel, H. and Avelino, P. P. and Avgoustidis, A. and Avila, S. and Awan, S. and Azzollini, R. and Baccigalupi, C. and Bachelet, E. and Bacon, D. and Baes, M. and Bagley, M. B. and Bahr-Kalus, B. and Balaguera-Antolinez, A. and Balbinot, E. and Balcells, M. and Baldi, M. and Baldry, I. and Balestra, A. and Ballardini, M. and Ballester, O. and Balogh, M. and Bañados, E. and Barbier, R. and Bardelli, S. and Baron, M. and Barreiro, T. and Barrena, R. and Barriere, J.-C. and Barros, B. J. and Barthelemy, A. and Bartolo, N. and Basset, A. and Battaglia, P. and Battisti, A. J. and Baugh, C. M. and Baumont, L. and Bazzanini, L. and Beaulieu, J.-P. and Beckmann, V. and Belikov, A. N. and Bel, J. and Bellagamba, F. and Bella, M. and Bellini, E. and Benabed, K. and Bender, R. and Benevento, G. and Bennett, C. L. and Benson, K. and Bergamini, P. and Bermejo-Climent, J. R. and Bernardeau, F. and Bertacca, D. and Berthe, M. and Berthier, J. and Bethermin, M. and Beutler, F. and Bevillon, C. and Bhargava, S. and Bhatawdekar, R. and Bianchi, D. and Bisigello, L. and Biviano, A. and Blake, R. P. and Blanchard, A. and Blazek, J. and Blot, L. and Bosco, A. and Bodendorf, C. and Boenke, T. and Böhringer, H. and Boldrini, P. and Bolzonella, M. and Bonchi, A. and Bonici, M. and Bonino, D. and Bonino, L. and Bonvin, C. and Bon, W. and Booth, J. T. and Borgani, S. and Borlaff, A. S. and Borsato, E. and Bosco, A. and Bose, B. and Botticella, M. T. and Boucaud, A. and Bouche, F. and Boucher, J. S. and Boutigny, D. and Bouvard, T. and Bouwens, R. and Bouy, H. and Bowler, R. A. A. and Bozza, V. and Bozzo, E. and Branchini, E. and Brando, G. and Brau-Nogue, S. and Brekke, P. and Bremer, M. N. and Brescia, M. and Breton, M.-A. and Brinchmann, J. and Brinckmann, T. and Brockley-Blatt, C. and Brodwin, M. and Brouard, L. and Brown, M. L. and Bruton, S. and Bucko, J. and Buddelmeijer, H. and Buenadicha, G. and Buitrago, F. and Burger, P. and Burigana, C. and Busillo, V. and Busonero, D. and Cabanac, R. and Cabayol-Garcia, L. and Cagliari, M. S. and Caillat, A. and Caillat, L. and Calabrese, M. and Calabro, A. and Calderone, G. and Calura, F. and Quevedo, B. Camacho and Camera, S. and Campos, L. and Canas-Herrera, G. and Candini, G. P. and Cantiello, M. and Capobianco, V. and Cappellaro, E. and Cappelluti, N. and Cappi, A. and Caputi, K. I. and Cara, C. and Carbone, C. and Cardone, V. F. and Carella, E. and Carlberg, R. G. and Carle, M. and Carminati, L. and Caro, F. and Carrasco, J. M. and Carretero, J. and Carrilho, P. and Duque, J. Carron and Carry, B. and Carvalho, A. and Carvalho, C. S. and Casas, R. and Casas, S. and Casenove, P. and Casey, C. M. and Cassata, P. and Castander, F. J. and Castelao, D. and Castellano, M. and Castiblanco, L. and Castignani, G. and Castro, T. and Cavet, C. and Cavuoti, S. and Chabaud, P.-Y. and Chambers, K. C. and Charles, Y. and Charlot, S. and Chartab, N. and Chary, R. and Chaumeil, F. and Cho, H. and Chon, G. and Ciancetta, E. and Ciliegi, P. and Cimatti, A. and Cimino, M. and Cioni, M.-R. L. and Claydon, R. and Cleland, C. and Clément, B. and Clements, D. L. and Clerc, N. and Clesse, S. and Codis, S. and Cogato, F. and Colbert, J. and Cole, R. E. and Coles, P. and Collett, T. E. and Collins, R. S. and Colodro-Conde, C. and Colombo, C. and Combes, F. and Conforti, V. and Congedo, G. and Conseil, S. and Conselice, C. J. and Contarini, S. and Contini, T. and Conversi, L. and Cooray, A. R. and Copin, Y. and Corasaniti, P.-S. and Corcho-Caballero, P. and Corcione, L. and Cordes, O. and Corpace, O. and Correnti, M. and Costanzi, M. and Costille, A. and Courbin, F. and Mifsud, L. Courcoult and Courtois, H. M. and Cousinou, M.-C. and Covone, G. and Cowell, T. and Cragg, C. and Cresci, G. and Cristiani, S. and Crocce, M. and Cropper, M. and Crouzet, P. E. and Csizi, B. and Cuby, J.-G. and Cucchetti, E. and Cucciati, O. and Cuillandre, J.-C. and Cunha, P. A. C. and Cuozzo, V. and Daddi, E. and D'Addona, M. and Dafonte, C. and Dagoneau, N. and Dalessandro, E. and Dalton, G. B. and D'Amico, G. and Dannerbauer, H. and Danto, P. and Das, I. and Silva, A. Da and Silva, R. da and Doumerg, W. d'Assignies and Daste, G. and Davies, J. E. and Davini, S. and Dayal, P. and Boer, T. de and Decarli, R. and Caro, B. De and Degaudenzi, H. and Degni, G. and Jong, J. T. A. de and Bella, L. F. de la and Torre, S. de la and Delhaise, F. and Delley, D. and Delucchi, G. and Lucia, G. De and Denniston, J. and Paolis, F. De and Petris, M. De and Derosa, A. and Desai, S. and Desjacques, V. and Despali, G. and Desprez, G. and Vicente-Albendea, J. De and Deville, Y. and Dias, J. D. F. and Díaz-Sánchez, A. and Diaz, J. J. and Domizio, S. Di and Diego, J. M. and Ferdinando, D. Di and Giorgio, A. M. Di and Dimauro, P. and Dinis, J. and Dolag, K. and Dolding, C. and Dole, H. and Sánchez, H. Domínguez and Doré, O. and Dournac, F. and Douspis, M. and Dreihahn, H. and Droge, B. and Dryer, B. and Dubath, F. and Duc, P.-A. and Ducret, F. and Duffy, C. and Dufresne, F. and Duncan, C. A. J. and Dupac, X. and Duret, V. and Durrer, R. and Durret, F. and Dusini, S. and Ealet, A. and Eggemeier, A. and Eisenhardt, P. R. M. and Elbaz, D. and Elkhashab, M. Y. and Ellien, A. and Endicott, J. and Enia, A. and Erben, T. and Vigo, J. A. Escartin and Escoffier, S. and Sanz, I. Escudero and Essert, J. and Ettori, S. and Ezziati, M. and Fabbian, G. and Fabricius, M. and Fang, Y. and Farina, A. and Farina, M. and Farinelli, R. and Farrens, S. and Faustini, F. and Feltre, A. and Ferguson, A. M. N. and Ferrando, P. and Ferrari, A. G. and Ferré-Mateu, A. and Ferreira, P. G. and Ferreras, I. and Ferrero, I. and Ferriol, S. and Ferruit, P. and Filleul, D. and Finelli, F. and Finkelstein, S. L. and Finoguenov, A. and Fiorini, B. and Flentge, F. and Focardi, P. and Fonseca, J. and Fontana, A. and Fontanot, F. and Fornari, F. and Fosalba, P. and Fossati, M. and Fotopoulou, S. and Fouchez, D. and Fourmanoit, N. and Frailis, M. and Fraix-Burnet, D. and Franceschi, E. and Franco, A. and Franzetti, P. and Freihoefer, J. and Frenk, C. S. and Frittoli, G. and Frugier, P.-A. and Frusciante, N. and Fumagalli, A. and Fumagalli, M. and Fumana, M. and Fu, Y. and Gabarra, L. and Galeotta, S. and Galluccio, L. and Ganga, K. and Gao, H. and García-Bellido, J. and Garcia, K. and Gardner, J. P. and Garilli, B. and Gaspar-Venancio, L.-M. and Gasparetto, T. and Gautard, V. and Gavazzi, R. and Gaztanaga, E. and Genolet, L. and Santos, R. Genova and Gentile, F. and George, K. and Gerbino, M. and Ghaffari, Z. and Giacomini, F. and Gianotti, F. and Gibb, G. P. S. and Gillard, W. and Gillis, B. and Ginolfi, M. and Giocoli, C. and Girardi, M. and Giri, S. K. and Goh, L. W. K. and Gómez-Alvarez, P. and Gonzalez-Perez, V. and Gonzalez, A. H. and Gonzalez, E. J. and Gonzalez, J. C. and Beauchamps, S. Gouyou and Gozaliasl, G. and Gracia-Carpio, J. and Grandis, S. and Granett, B. R. and Granvik, M. and Grazian, A. and Gregorio, A. and Grenet, C. and Grillo, C. and Grupp, F. and Gruppioni, C. and Gruppuso, A. and Guerbuez, C. and Guerrini, S. and Guidi, M. and Guillard, P. and Gutierrez, C. M. and Guttridge, P. and Guzzo, L. and Gwyn, S. and Haapala, J. and Haase, J. and Haddow, C. R. and Hailey, M. and Hall, A. and Hall, D. and Hamaus, N. and Haridasu, B. S. and Harnois-Déraps, J. and Harper, C. and Hartley, W. G. and Hasinger, G. and Hassani, F. and Hatch, N. A. and Haugan, S. V. H. and Häußler, B. and Heavens, A. and Heisenberg, L. and Helmi, A. and Helou, G. and Hemmati, S. and Henares, K. and Herent, O. and Hernández-Monteagudo, C. and Heuberger, T. and Hewett, P. C. and Heydenreich, S. and Hildebrandt, H. and Hirschmann, M. and Hjorth, J. and Hoar, J. and Hoekstra, H. and Holland, A. D. and Holliman, M. S. and Holmes, W. and Hook, I. and Horeau, B. and Hormuth, F. and Hornstrup, A. and Hosseini, S. and Hu, D. and Hudelot, P. and Hudson, M. J. and Huertas-Company, M. and Huff, E. M. and Hughes, A. C. N. and Humphrey, A. and Hunt, L. K. and Huynh, D. D. and Ibata, R. and Ichikawa, K. and Iglesias-Groth, S. and Ilbert, O. and Ilić, S. and Ingoglia, L. and Iodice, E. and Israel, H. and Israelsson, U. E. and Izzo, L. and Jablonka, P. and Jackson, N. and Jacobson, J. and Jafariyazani, M. and Jahnke, K. and Jain, B. and Jansen, H. and Jarvis, M. J. and Jasche, J. and Jauzac, M. and Jeffrey, N. and Jhabvala, M. and Jimenez-Teja, Y. and Muñoz, A. Jimenez and Joachimi, B. and Johansson, P. H. and Joudaki, S. and Jullo, E. and Kajava, J. J. E. and Kang, Y. and Kannawadi, A. and Kansal, V. and Karagiannis, D. and Kärcher, M. and Kashlinsky, A. and Kazandjian, M. V. and Keck, F. and Keihänen, E. and Kerins, E. and Kermiche, S. and Khalil, A. and Kiessling, A. and Kiiveri, K. and Kilbinger, M. and Kim, J. and King, R. and Kirkpatrick, C. C. and Kitching, T. and Kluge, M. and Knabenhans, M. and Knapen, J. H. and Knebe, A. and Kneib, J.-P. and Kohley, R. and Koopmans, L. V. E. and Koskinen, H. and Koulouridis, E. and Kou, R. and Kovács, A. and Kovačić, I. and Kowalczyk, A. and Koyama, K. and Kraljic, K. and Krause, O. and Kruk, S. and Kubik, B. and Kuchner, U. and Kuijken, K. and Kümmel, M. and Kunz, M. and Kurki-Suonio, H. and Lacasa, F. and Lacey, C. G. and Franca, F. La and Lagarde, N. and Lahav, O. and Laigle, C. and Marca, A. La and Marle, O. La and Lamine, B. and Lam, M. C. and Lançon, A. and Landt, H. and Langer, M. and Lapi, A. and Larcheveque, C. and Larsen, S. S. and Lattanzi, M. and Laudisio, F. and Laugier, D. and Laureijs, R. and Laurent, V. and Lavaux, G. and Lawrenson, A. and Lazanu, A. and Lazeyras, T. and Boulc'h, Q. Le and Brun, A. M. C. Le and Brun, V. Le and Leclercq, F. and Lee, S. and Graet, J. Le and Legrand, L. and Leirvik, K. N. and Jeune, M. Le and Lembo, M. and Mignant, D. Le and Lepinzan, M. D. and Lepori, F. and Reun, A. Le and Leroy, G. and Lesci, G. F. and Lesgourgues, J. and Leuzzi, L. and Levi, M. E. and Liaudat, T. I. and Libet, G. and Liebing, P. and Ligori, S. and Lilje, P. B. and Lin, C.-C. and Linde, D. and Linder, E. and Lindholm, V. and Linke, L. and Li, S.-S. and Liu, S. J. and Lloro, I. and Lobo, F. S. N. and Lodieu, N. and Lombardi, M. and Lombriser, L. and Lonare, P. and Longo, G. and López-Caniego, M. and Lopez, X. Lopez and Alvarez, J. Lorenzo and Loureiro, A. and Loveday, J. and Lusso, E. and Macias-Perez, J. and Maciaszek, T. and Maggio, G. and Magliocchetti, M. and Magnard, F. and Magnier, E. A. and Magro, A. and Mahler, G. and Mainetti, G. and Maino, D. and Maiorano, E. and Maiorano, E. and Malavasi, N. and Mamon, G. A. and Mancini, C. and Mandelbaum, R. and Manera, M. and Manjón-García, A. and Mannucci, F. and Mansutti, O. and Outeiro, M. Manteiga and Maoli, R. and Maraston, C. and Marcin, S. and Marcos-Arenal, P. and Margalef-Bentabol, B. and Marggraf, O. and Marinucci, D. and Marinucci, M. and Markovic, K. and Marleau, F. R. and Marpaud, J. and Martignac, J. and Martín-Fleitas, J. and Martin-Moruno, P. and Martin, E. L. and Martinelli, M. and Martinet, N. and Martin, H. and Martins, C. J. A. P. and Marulli, F. and Massari, D. and Massey, R. and Masters, D. C. and Matarrese, S. and Matsuoka, Y. and Matthew, S. and Maughan, B. J. and Mauri, N. and Maurin, L. and Maurogordato, S. and McCarthy, K. and McConnachie, A. W. and McCracken, H. J. and McDonald, I. and McEwen, J. D. and McPartland, C. J. R. and Medinaceli, E. and Mehta, V. and Mei, S. and Melchior, M. and Melin, J.-B. and Ménard, B. and Mendes, J. and Mendez-Abreu, J. and Meneghetti, M. and Mercurio, A. and Merlin, E. and Metcalf, R. B. and Meylan, G. and Migliaccio, M. and Mignoli, M. and Miller, L. and Miluzio, M. and Milvang-Jensen, B. and Mimoso, J. P. and Miquel, R. and Miyatake, H. and Mobasher, B. and Mohr, J. J. and Monaco, P. and Monguió, M. and Montoro, A. and Mora, A. and Dizgah, A. Moradinezhad and Moresco, M. and Moretti, C. and Morgante, G. and Morisset, N. and Moriya, T. J. and Morris, P. W. and Mortlock, D. J. and Moscardini, L. and Mota, D. F. and Mottet, S. and Moustakas, L. A. and Moutard, T. and Müller, T. and Munari, E. and Murphree, G. and Murray, C. and Murray, N. and Musi, P. and Nadathur, S. and Nagam, B. C. and Nagao, T. and Naidoo, K. and Nakajima, R. and Nally, C. and Natoli, P. and Navarro-Alsina, A. and Girones, D. Navarro and Neissner, C. and Nersesian, A. and Nesseris, S. and Nguyen-Kim, H. N. and Nicastro, L. and Nichol, R. C. and Nielbock, M. and Niemi, S.-M. and Nieto, S. and Nilsson, K. and Noller, J. and Norberg, P. and Nouri-Zonoz, A. and Ntelis, P. and Nucita, A. A. and Nugent, P. and Nunes, N. J. and Nutma, T. and Ocampo, I. and Odier, J. and Oesch, P. A. and Oguri, M. and Oliveira, D. Magalhaes and Onoue, M. and Oosterbroek, T. and Oppizzi, F. and Ordenovic, C. and Osato, K. and Pacaud, F. and Pace, F. and Padilla, C. and Paech, K. and Pagano, L. and Page, M. J. and Palazzi, E. and Paltani, S. and Pamuk, S. and Pandolfi, S. and Paoletti, D. and Paolillo, M. and Papaderos, P. and Pardede, K. and Parimbelli, G. and Parmar, A. and Partmann, C. and Pasian, F. and Passalacqua, F. and Paterson, K. and Patrizii, L. and Pattison, C. and Paulino-Afonso, A. and Paviot, R. and Peacock, J. A. and Pearce, F. R. and Pedersen, K. and Peel, A. and Peletier, R. F. and Ibanez, M. Pellejero and Pello, R. and Penny, M. T. and Percival, W. J. and Perez-Garrido, A. and Perotto, L. and Pettorino, V. and Pezzotta, A. and Pezzuto, S. and Philippon, A. and Pierre, M. and Piersanti, O. and Pietroni, M. and Piga, L. and Pilo, L. and Pires, S. and Pisani, A. and Pizzella, A. and Pizzuti, L. and Plana, C. and Polenta, G. and Pollack, J. E. and Poncet, M. and Pöntinen, M. and Pool, P. and Popa, L. A. and Popa, V. and Popp, J. and Porciani, C. and Porth, L. and Potter, D. and Poulain, M. and Pourtsidou, A. and Pozzetti, L. and Prandoni, I. and Pratt, G. W. and Prezelus, S. and Prieto, E. and Pugno, A. and Quai, S. and Quilley, L. and Racca, G. D. and Raccanelli, A. and Rácz, G. and Radinović, S. and Radovich, M. and Ragagnin, A. and Ragnit, U. and Raison, F. and Ramos-Chernenko, N. and Ranc, C. and Rasera, Y. and Raylet, N. and Rebolo, R. and Refregier, A. and Reimberg, P. and Reiprich, T. H. and Renk, F. and Renzi, A. and Retre, J. and Revaz, Y. and Reylé, C. and Reynolds, L. and Rhodes, J. and Ricci, F. and Ricci, M. and Riccio, G. and Ricken, S. O. and Rissanen, S. and Risso, I. and Rix, H.-W. and Robin, A. C. and Rocca-Volmerange, B. and Rocci, P.-F. and Rodenhuis, M. and Rodighiero, G. and Monroy, M. Rodriguez and Rollins, R. P. and Romanello, M. and Roman, J. and Romelli, E. and Romero-Gomez, M. and Roncarelli, M. and Rosati, P. and Rosset, C. and Rossetti, E. and Roster, W. and Rottgering, H. J. A. and Rozas-Fernández, A. and Ruane, K. and Rubino-Martin, J. A. and Rudolph, A. and Ruppin, F. and Rusholme, B. and Sacquegna, S. and Sáez-Casares, I. and Saga, S. and Saglia, R. and Sahlén, M. and Saifollahi, T. and Sakr, Z. and Salvalaggio, J. and Salvaterra, R. and Salvati, L. and Salvato, M. and Salvignol, J.-C. and Sánchez, A. G. and Sanchez, E. and Sanders, D. B. and Sapone, D. and Saponara, M. and Sarpa, E. and Sarron, F. and Sartori, S. and Sartoris, B. and Sassolas, B. and Sauniere, L. and Sauvage, M. and Sawicki, M. and Scaramella, R. and Scarlata, C. and Scharré, L. and Schaye, J. and Schewtschenko, J. A. and Schindler, J.-T. and Schinnerer, E. and Schirmer, M. and Schmidt, F. and Schmidt, F. and Schmidt, M. and Schneider, A. and Schneider, M. and Schneider, P. and Schöneberg, N. and Schrabback, T. and Schultheis, M. and Schulz, S. and Schuster, N. and Schwartz, J. and Sciotti, D. and Scodeggio, M. and Scognamiglio, D. and Scott, D. and Scottez, V. and Secroun, A. and Sefusatti, E. and Seidel, G. and Seiffert, M. and Sellentin, E. and Selwood, M. and Semboloni, E. and Sereno, M. and Serjeant, S. and Serrano, S. and Setnikar, G. and Shankar, F. and Sharples, R. M. and Short, A. and Shulevski, A. and Shuntov, M. and Sias, M. and Sikkema, G. and Silvestri, A. and Simon, P. and Sirignano, C. and Sirri, G. and Skottfelt, J. and Slezak, E. and Sluse, D. and Smith, G. P. and Smith, L. C. and Smith, R. E. and Smit, S. J. A. and Soldano, F. and Solheim, B. G. B. and Sorce, J. G. and Sorrenti, F. and Soubrie, E. and Spinoglio, L. and Mancini, A. Spurio and Stadel, J. and Stagnaro, L. and Stanco, L. and Stanford, S. A. and Starck, J.-L. and Stassi, P. and Steinwagner, J. and Stern, D. and Stone, C. and Strada, P. and Strafella, F. and Stramaccioni, D. and Surace, C. and Sureau, F. and Suyu, S. H. and Swindells, I. and Szafraniec, M. and Szapudi, I. and Taamoli, S. and Talia, M. and Tallada-Crespí, P. and Tanidis, K. and Tao, C. and Tarrío, P. and Tavagnacco, D. and Taylor, A. N. and Taylor, J. E. and Taylor, P. L. and Teixeira, E. M. and Tenti, M. and Idiago, P. Teodoro and Teplitz, H. I. and Tereno, I. and Tessore, N. and Testa, V. and Testera, G. and Tewes, M. and Teyssier, R. and Theret, N. and Thizy, C. and Thomas, P. D. and Toba, Y. and Toft, S. and Toledo-Moreo, R. and Tolstoy, E. and Tommasi, E. and Torbaniuk, O. and Torradeflot, F. and Tortora, C. and Tosi, S. and Tosti, S. and Trifoglio, M. and Troja, A. and Trombetti, T. and Tronconi, A. and Tsedrik, M. and Tsyganov, A. and Tucci, M. and Tutusaus, I. and Uhlemann, C. and Ulivi, L. and Urbano, M. and Vacher, L. and Vaillon, L. and Valageas, P. and Valdes, I. and Valentijn, E. A. and Valenziano, L. and Valieri, C. and Valiviita, J. and Broeck, M. Van den and Vassallo, T. and Vavrek, R. and Vega-Ferrero, J. and Venemans, B. and Venhola, A. and Ventura, S. and Kleijn, G. Verdoes and Vergani, D. and Verma, A. and Vernizzi, F. and Veropalumbo, A. and Verza, G. and Vescovi, C. and Vibert, D. and Viel, M. and Vielzeuf, P. and Viglione, C. and Viitanen, A. and Villaescusa-Navarro, F. and Vinciguerra, S. and Visticot, F. and Voggel, K. and Wietersheim-Kramsta, M. von and Vriend, W. J. and Wachter, S. and Walmsley, M. and Walth, G. and Walton, D. M. and Walton, N. A. and Wander, M. and Wang, L. and Wang, Y. and Weaver, J. R. and Weller, J. and Wetzstein, M. and Whalen, D. J. and Whittam, I. H. and Widmer, A. and Wiesmann, M. and Wilde, J. and Williams, O. R. and Winther, H.-A. and Wittje, A. and Wong, J. H. W. and Wright, A. H. and Yankelevich, V. and Yeung, H. W. and Yoon, M. and Youles, S. and Yung, L. Y. A. and Zacchei, A. and Zalesky, L. and Zamorani, G. and Vitorelli, A. Zamorano and Marc, M. Zanoni and Zennaro, M. and Zerbi, F. M. and Zinchenko, I. A. and Zoubian, J. and Zucca, E. and Zumalacarregui, M.},
	month = may,
	year = {2025},
	note = {arXiv:2405.13491 [astro-ph.CO]},
	pages = {A1},
}

@article{abareshi_overview_2022,
	title = {Overview of the {Instrumentation} for the {Dark} {Energy} {Spectroscopic} {Instrument}},
	volume = {164},
	issn = {0004-6256, 1538-3881},
	url = {http://arxiv.org/abs/2205.10939},
	doi = {10.3847/1538-3881/ac882b},
	number = {5},
	urldate = {2025-11-04},
	journal = {The Astronomical Journal},
	author = {Abareshi, B. and Aguilar, J. and Ahlen, S. and Alam, Shadab and Alexander, David M. and Alfarsy, R. and Allen, L. and Prieto, C. Allende and Alves, O. and Ameel, J. and Armengaud, E. and Asorey, J. and Aviles, Alejandro and Bailey, S. and Balaguera-Antolínez, A. and Ballester, O. and Baltay, C. and Bault, A. and Beltran, S. F. and Benavides, B. and BenZvi, S. and Berti, A. and Besuner, R. and Beutler, Florian and Bianchi, D. and Blake, C. and Blanc, P. and Blum, R. and Bolton, A. and Bose, S. and Bramall, D. and Brieden, S. and Brodzeller, A. and Brooks, D. and Brownewell, C. and Buckley-Geer, E. and Cahn, R. N. and Cai, Z. and Canning, R. and Rosell, A. Carnero and Carton, P. and Casas, R. and Castander, F. J. and Cervantes-Cota, J. L. and Chabanier, S. and Chaussidon, E. and Chuang, C. and Circosta, C. and Cole, S. and Cooper, A. P. and Costa, L. da and Cousinou, M.-C. and Cuceu, A. and Davis, T. M. and Dawson, K. and Cruz-Noriega, R. de la and Macorra, A. de la and Mattia, A. de and Costa, J. Della and Demmer, P. and Derwent, M. and Dey, A. and Dey, B. and Dhungana, G. and Ding, Z. and Dobson, C. and Doel, P. and Donald-McCann, J. and Donaldson, J. and Douglass, K. and Duan, Y. and Dunlop, P. and Edelstein, J. and Eftekharzadeh, S. and Eisenstein, D. J. and Enriquez-Vargas, M. and Escoffier, S. and Evatt, M. and Fagrelius, P. and Fan, X. and Fanning, K. and Fawcett, V. A. and Ferraro, S. and Ereza, J. and Flaugher, B. and Font-Ribera, A. and Forero-Romero, J. E. and Frenk, C. S. and Fromenteau, S. and Gänsicke, B. T. and Garcia-Quintero, C. and Garrison, L. and Gaztañaga, E. and Gerardi, F. and Gil-Marín, H. and Gontcho, S. Gontcho A. and Gonzalez-Morales, Alma X. and Gonzalez-de-Rivera, G. and Gonzalez-Perez, V. and Gordon, C. and Graur, O. and Green, D. and Grove, C. and Gruen, D. and Gutierrez, G. and Guy, J. and Hahn, C. and Harris, S. and Herrera, D. and Herrera-Alcantar, Hiram K. and Honscheid, K. and Howlett, C. and Huterer, D. and Iršič, V. and Ishak, M. and Jelinsky, P. and Jiang, L. and Jimenez, J. and Jing, Y. P. and Joyce, R. and Jullo, E. and Juneau, S. and Karaçaylı, N. G. and Karamanis, M. and Karcher, A. and Karim, T. and Kehoe, R. and Kent, S. and Kirkby, D. and Kisner, T. and Kitaura, F. and Koposov, S. E. and Kovács, A. and Kremin, A. and Krolewski, Alex and L'Huillier, B. and Lahav, O. and Lambert, A. and Lamman, C. and Lan, Ting-Wen and Landriau, M. and Lane, S. and Lang, D. and Lange, J. U. and Lasker, J. and Guillou, L. Le and Leauthaud, A. and Suu, A. Le Van and Levi, Michael E. and Li, T. S. and Magneville, C. and Manera, M. and Manser, Christopher J. and Marshall, B. and McCollam, W. and McDonald, P. and Meisner, Aaron M. and Mezcua, J. Mena-Fernández M. and Miller, T. and Miquel, R. and Montero-Camacho, P. and Moon, J. and Martini, J. Paul and Meneses-Rizo, J. and Moustakas, J. and Mueller, E. and Muñoz-Gutiérrez, Andrea and Myers, Adam D. and Nadathur, S. and Najita, J. and Napolitano, L. and Neilsen, E. and Newman, Jeffrey A. and Nie, J. D. and Ning, Y. and Niz, G. and Norberg, P. and Noriega, Hernán E. and O'Brien, T. and Obuljen, A. and Palanque-Delabrouille, N. and Palmese, A. and Zhiwei, P. and Pappalardo, D. and Peng, X. and Percival, W. J. and Perruchot, S. and Pogge, R. and Poppett, C. and Porredon, A. and Prada, F. and Prochaska, J. and Pucha, R. and Pérez-Fernández, A. and Pérez-Ráfols, I. and Rabinowitz, D. and Raichoor, A. and Ramirez-Solano, S. and Ramírez-Pérez, César and Ravoux, C. and Reil, K. and Rezaie, M. and Rocher, A. and Rockosi, C. and Roe, N. A. and Roodman, A. and Ross, A. J. and Rossi, G. and Ruggeri, R. and Ruhlmann-Kleider, V. and Sabiu, C. G. and Safonova, S. and Said, K. and Saintonge, A. and Catonga, Javier Salas and Samushia, L. and Sanchez, E. and Saulder, C. and Schaan, E. and Schlafly, E. and Schlegel, D. and Schmoll, J. and Scholte, D. and Schubnell, M. and Secroun, A. and Seo, H. and Serrano, S. and Sharples, Ray M. and Sholl, Michael J. and Silber, Joseph Harry and Silva, D. R. and Sirk, M. and Siudek, M. and Smith, A. and Sprayberry, D. and Staten, R. and Stupak, B. and Tan, T. and Tarlé, Gregory and Tie, Suk Sien and Tojeiro, R. and Ureña-López, L. A. and Valdes, F. and Valenzuela, O. and Valluri, M. and Vargas-Magaña, M. and Verde, L. and Walther, M. and Wang, B. and Wang, M. S. and Weaver, B. A. and Weaverdyck, C. and Wechsler, R. and Wilson, Michael J. and Yang, J. and Yu, Y. and Yuan, S. and Yèche, Christophe and Zhang, H. and Zhang, K. and Zhao, Cheng and Zhou, Rongpu and Zhou, Zhimin and Zou, H. and Zou, J. and Zou, S. and Zu, Y.},
	month = nov,
	year = {2022},
	note = {arXiv:2205.10939 [astro-ph]},
	pages = {207},
}

@article{bower_breaking_2006,
	title = {Breaking the hierarchy of galaxy formation},
	volume = {370},
	issn = {0035-8711},
	url = {https://ui.adsabs.harvard.edu/abs/2006MNRAS.370..645B},
	doi = {10.1111/j.1365-2966.2006.10519.x},
	urldate = {2026-09-16},
	journal = {Monthly Notices of the Royal Astronomical Society},
	publisher = {OUP},
	author = {Bower, R. G. and Benson, A. J. and Malbon, R. and Helly, J. C. and Frenk, C. S. and Baugh, C. M. and Cole, S. and Lacey, C. G.},
	month = aug,
	year = {2006},
	note = {ADS Bibcode: 2006MNRAS.370..645B},
	pages = {645--655},
}

@misc{chudaykin_reanalyzing_2025,
	title = {Reanalyzing {DESI} {DR1}: 2. {Constraints} on {Dark} {Energy}, {Spatial} {Curvature}, and {Neutrino} {Masses}},
	shorttitle = {Reanalyzing {DESI} {DR1}},
	url = {http://arxiv.org/abs/2511.20757},
	doi = {10.48550/arXiv.2511.20757},
	urldate = {2026-03-23},
	publisher = {arXiv},
	author = {Chudaykin, Anton and Ivanov, Mikhail M. and Philcox, Oliver H. E.},
	month = nov,
	year = {2025},
	note = {arXiv:2511.20757 [astro-ph]},
}
\end{document}